\documentclass{emulateapj}
\shortauthors{Matthews et al.}
\shorttitle{The ALMA Phasing System}

\begin{document}
\newcommand{\ang}{\rm \AA}
\newcommand{\msun}{M$_\odot$}
\newcommand{\lsun}{L$_\odot$}
\newcommand{\days}{$d$}
\newcommand{\degree}{$^\circ$}
\newcommand{\ud}{{\rm d}}
\newcommand{\as}[2]{$#1''\,\hspace{-1.7mm}.\hspace{.0mm}#2$}
\newcommand{\am}[2]{$#1'\,\hspace{-1.7mm}.\hspace{.0mm}#2$}
\newcommand{\ad}[2]{$#1^{\circ}\,\hspace{-1.7mm}.\hspace{.0mm}#2$}
\newcommand{\lsim}{~\rlap{$<$}{\lower 1.0ex\hbox{$\sim$}}}
\newcommand{\gsim}{~\rlap{$>$}{\lower 1.0ex\hbox{$\sim$}}}
\newcommand{\HA}{H$\alpha$}
\newcommand{\HII}{\mbox{H\,{\sc ii}}}
\newcommand{\kms}{\mbox{km s$^{-1}$}}
\newcommand{\HI}{\mbox{H\,{\sc i}}}
\newcommand{\jks}{Jy~km~s$^{-1}$}

\title{The ALMA Phasing System: A Beamforming Capability for 
Ultra-High-Resolution Science at (Sub)Millimeter Wavelengths}

\author{
L. D. Matthews\altaffilmark{1},  
G. B. Crew\altaffilmark{1}, 
S. S. Doeleman\altaffilmark{1,2}, 
R. Lacasse\altaffilmark{3},
A. F. Saez\altaffilmark{4}, 
W. Alef\altaffilmark{5},
K. Akiyama\altaffilmark{1,6},
R. Amestica\altaffilmark{3}, 
J. M. Anderson\altaffilmark{5,7},
D. A. Barkats\altaffilmark{2,4}, 
A. Baudry\altaffilmark{8},
D. Brogui\`ere\altaffilmark{9},
R. Escoffier\altaffilmark{3},
V. L. Fish\altaffilmark{1}, 
J. Greenberg\altaffilmark{3},  
M. H. Hecht\altaffilmark{1},
R. Hiriart\altaffilmark{10}, 
A. Hirota\altaffilmark{4}, 
M. Honma\altaffilmark{6,11},
P. T. P. Ho\altaffilmark{12}, 
C. M. V. Impellizzeri\altaffilmark{4}, 
M. Inoue\altaffilmark{12}, 
Y. Kohno\altaffilmark{6},
B. Lopez\altaffilmark{4}, 
I. Mart\'\i-Vidal\altaffilmark{13}, 
H. Messias\altaffilmark{4,14}, 
Z. Meyer-Zhao\altaffilmark{12}, 
M. Mora-Klein\altaffilmark{3}, 
N. M. Nagar\altaffilmark{15}, 
H. Nishioka\altaffilmark{12}, 
T. Oyama\altaffilmark{6},
V. Pankratius\altaffilmark{1}, 
J. Perez\altaffilmark{3}, 
N. Phillips\altaffilmark{4}, 
N. Pradel\altaffilmark{12}, 
H. Rottmann\altaffilmark{5}, 
A. L. Roy\altaffilmark{5}, 
C. A. Ruszczyk\altaffilmark{1},
B. Shillue\altaffilmark{3}, 
S. Suzuki\altaffilmark{6},
R. Treacy\altaffilmark{3}
}
\altaffiltext{1}{Massachusetts Institute of Technology Haystack Observatory, 99 Millstone Road, Westford, MA
  01886 USA}
\altaffiltext{2}{Harvard-Smithsonian Center for Astrophysics, 60
  Garden Street, Cambridge, MA 02138 USA}
\altaffiltext{3}{National Radio Astronomy Observatory, NRAO Technology
 Center, 1180 Boxwood Estate Road, Charlottesville, VA 22903 USA}
\altaffiltext{4}{Joint ALMA Observatory, Alonso de C\'ordova 3107,
  Vitacura 763-0355, Santiago de Chile, Chile}
\altaffiltext{5}{Max-Planck-Institut f\"ur Radioastronomie, Auf dem
  H\"ugel 69, 53121 Bonn, Germany}
\altaffiltext{6}{Mizusawa VLBI Observatory, National Astronomical
  Observatory of Japan, Ohshu, Iwate 023-0861, Japan}
\altaffiltext{7}{Deutsches GeoForschungsZentrum, Telegrafenberg, 14473 Potsdam, Germany}
\altaffiltext{8}{Univ. Bordeaux, LAB, B18N, all\'ee Geoffroy Saint-Hilaire, F-33615 Pessac, France}
\altaffiltext{9}{Institut de Radio Astronomie Millim\'etrique, 300
  rue de la Piscine, Domaine Universitaire, F-38406 Saint Martin d'H\'eres,
 France}
\altaffiltext{10}{National Radio Astronomy Observatory, PO Box O,
  Socorro, NM 87801 USA} 
\altaffiltext{11}{SOKENDAI (The Graduate University for the Advanced
  Studies), Mitaka, Tokyo 181-8588, Japan}
\altaffiltext{12}{Institute of Astronomy and Astrophysics, Academia
  Sinica, P. O. Box 23-141, Taipei 10617, Taiwan}
\altaffiltext{13}{Department of Space, Earth, and Environment, Onsala
  Space Observatory, Chalmers
  University of Technology,  43992 Onsala,
  Sweden}
\altaffiltext{14}{European Southern Observatory, Alonso de C\'ordova
  3107, Vitacura, 
Casilla 19001, Santiago de Chile, Chile}
\altaffiltext{15}{Universidad de Concepci\'on, V\'\i ctor Lamas 1290,
  Casilla 160-C, Concepci\'on, Chile}

\begin{abstract}
The Atacama Millimeter/submillimeter Array (ALMA) Phasing Project
(APP) has developed and deployed the hardware and software 
necessary to coherently sum the signals of
individual ALMA antennas and record the aggregate sum in Very Long
Baseline Interferometry (VLBI) Data
Exchange Format. These beamforming
capabilities allow the ALMA array to collectively 
function as the equivalent of a single large aperture and participate
in global VLBI arrays. The inclusion of
phased ALMA in current VLBI networks  operating at (sub)millimeter wavelengths
provides an order of magnitude 
improvement in sensitivity, as well as enhancements in
$u$-$v$ coverage and north-south angular resolution. 
The availability of a phased ALMA enables a wide range of new ultra-high 
angular resolution 
science applications, including the resolution of supermassive black
holes on event horizon scales and studies of the launch and
collimation of astrophysical jets. It also provides a high-sensitivity
aperture that may be used for investigations such as pulsar searches at high frequencies.
This paper 
provides an overview of the ALMA Phasing System
design, implementation, and performance characteristics.  
\end{abstract}

\keywords{instrumentation: high angular resolution -- instrumentation:
  interferometers -- methods: observational}  

\section{Background\protect\label{background}}
The technique of Very Long Baseline Interferometry (VLBI) provides
measurements of astronomical sources with the highest angular
resolution presently achievable. At centimeter wavelengths, VLBI has
been used since the 1960s
to measure the sizes and structures of radio sources on scales finer
than a milliarcsecond (see Kellermann \&
Cohen 1988; Walker 1999). Extension of VLBI techniques to the
millimeter (mm) regime brings into reach still
higher angular resolution---as fine as a few tens of
microarcseconds. VLBI at mm wavelengths is also uniquely powerful in
enabling the study of sources 
that are scatter-broadened or self-absorbed at longer
wavelengths (e.g., extragalactic radio sources and 
supermassive black holes; Krichbaum 2003; Doeleman et al. 2008; Hada
et al. 2013; Boccardi et al. 2016), or
that emit high brightness temperature emission from spectral lines in
mm bands (e.g., Doeleman et al. 2002, 2004; Richter, Kemball, \& Jonas
2016; Issaoun et al. 2017). 

The successful expansion of VLBI into the mm domain during the 1980s required both technological
advances and the development of special algorithms to overcome the
challenges posed by the decreasing atmospheric coherence
timescales at shorter wavelengths (Rogers et al. 1984, 1995; Johnson
et al. 2015).
The first successful VLBI experiments at wavelengths as short as $\sim$1~mm
involved only a single baseline (Padin et al.\ 1990; Greve et
al.\ 1995; Krichbaum et al.\ 1997).  Various experiments over the 
decade that followed met with mixed success owing to the small bandwidths that could
be observed with the analog systems of the era.  The development and
deployment of digital backends (DBEs) and new recording systems based on hard
drives (namely the Mark~5 recording systems; Whitney et al. 2010) 
provided a much-needed boost in sensitivity.  These and other
technical developments culminated in 2007 with the first detection of
Schwarzschild radius--scale structure in Sagittarius~A* (Sgr~A*)  at
1.3~mm on baselines
from the Arizona Radio Observatory's Submillimeter Telescope to the
James Clerk Maxwell Telescope and a telescope of the Combined Array
for Research in Millimeter-wave Astronomy (CARMA; Doeleman
et al.\ 2008).  Motivated by this success, the 1.3~mm observing array,
known as the Event Horizon Telescope (EHT; Doeleman et al. 2009), was expanded over the
following years to include the Submillimeter Array (SMA), the Institut de
Radioastronomie Millim\'{e}trique (IRAM) 30~m telescope, the Atacama
Pathfinder EXperiment (APEX), the Large Millimeter Telescope Alfonso
Serrano (LMT),
and the South Pole Telescope (SPT), along with the now-defunct Caltech
Submillimeter Observatory.  Key results from these observations
included the detection of ordered magnetic fields in an asymmetrically
bright emission region around Sgr~A* (Johnson et al.\ 2015; Fish et
al.\ 2016) and very compact structure at the base of the jet from the
supermassive black hole in M87 (Doeleman et al.\ 2012).

Despite these successes, the use of mm VLBI techniques has been restricted to
the study of a relatively small number of bright sources owing to the
limited sensitivity of existing networks of VLBI antennas. This is a
consequence both of  higher
receiver noise and the relatively small apertures (or effective
apertures) of most  telescopes operating at frequencies of
$\gsim$100~GHz. Furthermore, at wavelengths shorter than 3~mm, the small number of suitably
equipped VLBI stations have offered too few baselines
to allow sources to be imaged (e.g., Doeleman et al. 2009).  

As a major step toward overcoming these limitations, the ALMA Phasing
Project (APP) was conceived with the goal of harnessing 
the extraordinary sensitivity
of the Atacama Large Millimeter/submillimeter Array (ALMA) for VLBI
science. While capturing data from each separate ALMA antenna
is neither practical nor desirable, ALMA can be used as a single
large-aperture VLBI antenna if the data from its individual antennas are
phase-corrected and coherently added together. The goals of the APP
were to provide the hardware and software necessary to perform these
operations and to record the resulting data in VLBI format.

Phased arrays have been used as VLBI stations at centimeter
wavelengths since the late
1970s, beginning with the Westerbork Synthesis Radio Telescope (e.g.,
Kapahi \& Schilizzi 1979; Schilizzi \& Gurvits 1996; Raimond 1996),
and also including the phased Very Large Array (VLA; e.g., Wrobel 1983;
Lo et al. 1985) and the phased Australia Telescope Compact Array (e.g.,
Tzioumis 1997).  More recently, 
phased arrays, including the Plateau de Bure\footnote{{\url
    http://www.iram.fr/IRAMFR/TA/backend/cor6A/index.html}}, 
SMA, and CARMA, have also been deployed
for VLBI in the mm regime
(e.g., Krichbaum et al. 2008; 
Weintroub 2008).  Some of the factors affecting the performance and
efficiency of these earlier phased
arrays have been discussed by, e.g., van Ardenne (1979, 1980); Ulvestad (1988); Dewey (1994); Moran
(1989); and Kokkeler, Fridman, \& van Ardenne (2001). 

\begin{figure}
\center{\fbox{%
\includegraphics[width=3.0in,angle=0]{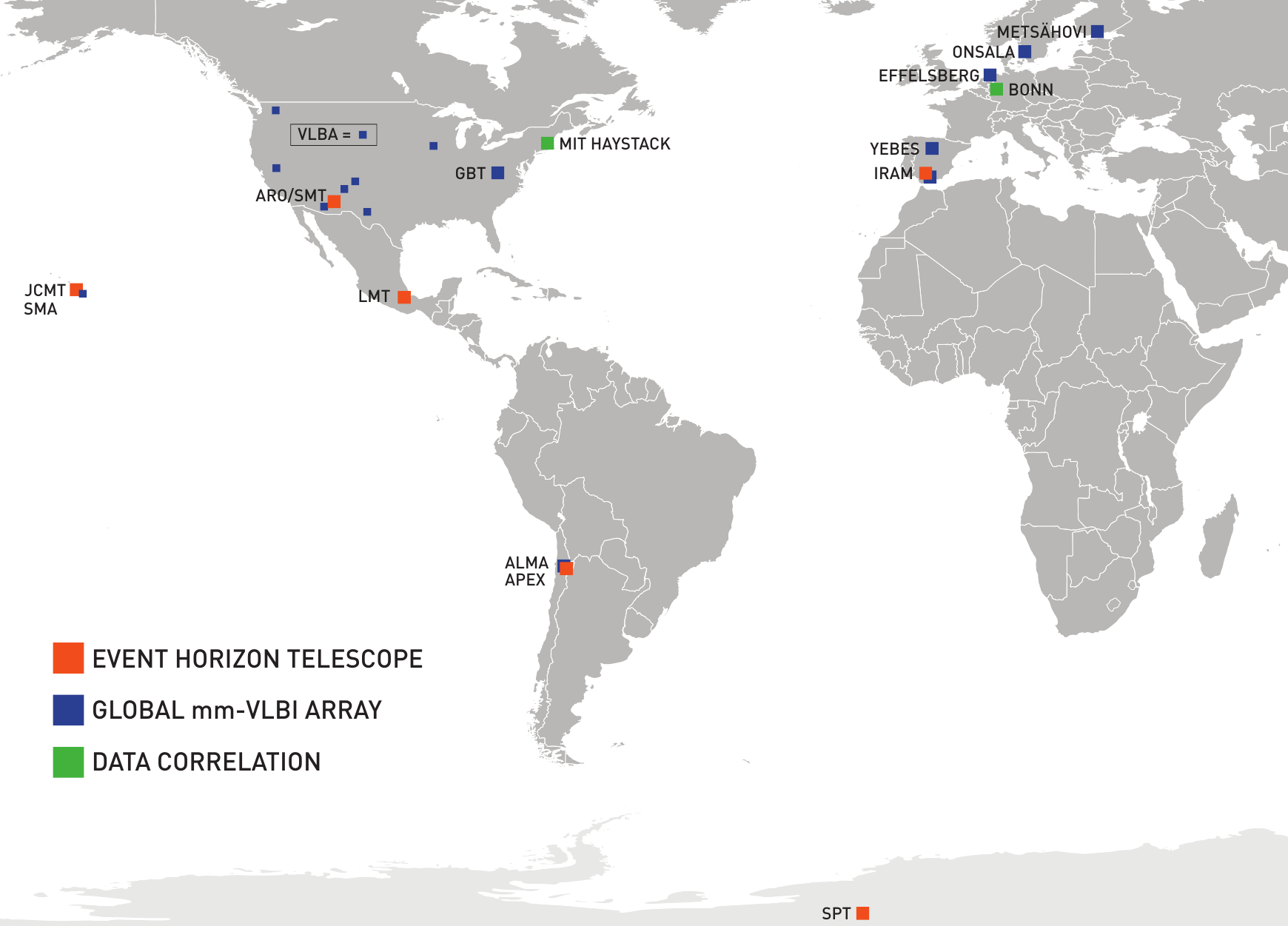}}}
\caption[Map]{
Map showing the global distribution of stations operating in
conjunction with phased ALMA for VLBI during ALMA Cycles 4 and
5. Sites with
Band~6 (1~mm) capabilities (part of the EHT) are shown in red; stations with
Band~3 (3~mm) capabilities
(part of the GMVA) are shown in blue; correlation centers (see Section~\ref{sec:korrel})
are shown in green. Sites that are part of the VLBA are not
individually labeled and are
distinguished from the other GMVA sites by a smaller symbol.}
\label{fig:map}
\end{figure}

An optimally phased array provides a
collecting area equivalent to the combined effective area of the
individual antennas, thereby boosting the achievable signal-to-noise ratio
(SNR) of  VLBI baselines to the site. The addition of a phased 
ALMA to existing VLBI networks operating at
mm wavelengths therefore has the ability to offer up to an order of magnitude boost in
sensitivity. For many experiments,  
the geographical location of ALMA also provides a significant improvement in $u$-$v$
coverage, dramatically enhancing
the ability to reconstruct images of sources.  And when  used in
conjunction with the Global mm-VLBI Array
(GMVA) at 3~mm, ALMA will also provide a factor of two improvement in north-south angular
resolution. Figure~\ref{fig:map} illustrates the geographic
distribution of VLBI antennas currently available to operate in
conjunction with ALMA at wavelengths of 1~mm and/or 3~mm. The 1~mm
observing sites are part of the EHT, as described above. The GMVA
sites include eight stations of the Very Long Baseline Array (VLBA),
along with
the Robert C. Byrd Green Bank Telescope (GBT), the Effelsberg 100~m
Radio Telescope, the Yebes Observatory 40~m telescope, the
IRAM 30~m telescope, the Mets\"ahovi 14~m telescope, and the Onsala
Space Observatory 20~m telescope.

The science case for a beamformed ALMA, described in detail by Fish et
al.\ (2013) and Tilanus et al.\ (2014), is broad and diverse.  VLBI
observations of supermassive black holes on event-horizon scales can
be used to gain a better understanding of the astrophysics of
accretion in black hole systems and to test predictions of General
Relativity.  High-resolution imaging of the jet launching, acceleration, and collimation
regions in active galactic nuclei will illuminate how jets are
formed, and exquisitely detailed studies of the nearest extragalactic jets, such as
the one in M87, will provide a benchmark for the general understanding of jet
physics (e.g., Asada \& Nakamura 2012; Asada et al. 2014).

High-frequency pulsar observations enabled by a phased ALMA (see also
Section~\ref{future}) will aid our understanding of
pulsar emission processes, and pulsar searches in the Galactic Center
may uncover a system that can be used for relativistic tests at very
high precision.  (Sub)mm maser observations at VLBI resolutions 
can be used to better understand
the physical conditions in star-forming regions, evolved stars 
(on AU scales), and the circumnuclear
environment of other galaxies, and multi-epoch observations of maser sources
have the potential to provide higher astrometric
accuracy than has been possible previously.  In addition, new observations of absorbing systems at
cosmological distances are poised to provide information on the physical conditions
of the early universe and chemical evolution over time and can also be
used to test for variability of fundamental physical constants.

The initial implementation of the ALMA Phasing System (APS) focused on the
simplest use case: continuum VLBI on sources bright enough to allow
on-source phasing of the array.  
To date, the APS has been commissioned and
approved for science observations in Bands 3 and 6 (corresponding to
wavelengths of 3~mm and 1.3~mm, respectively), where 
the GMVA and the EHT, respectively 
are available to serve as the respective partner networks (Figure~\ref{fig:map}). The first
science observations that included the APS in this capacity
were conducted in 2017
April as part of ALMA Cycle~4. 

The outline of this paper is as follows.
We first briefly summarize some of the attributes of the ALMA array
relevant to its use as a VLBI station (Section~\ref{ALMArocks}).
In Section~\ref{overview}, we provide an overview of the APS and
subsequently describe in more detail the hardware
(Section~\ref{hardware}) and software components (Section~\ref{sec:SW})
developed by the APP team to enable operation of ALMA as a phased array and a VLBI
station, as well as the specialized procedures required for
correlation of the resulting data (Sections~\ref{sec:korrel} and
\ref{polconvert}). 
In Section~\ref{CSV} we describe some of the
performance characteristics of phased ALMA derived from on-sky commissioning
and science verification. Finally, in
Section~\ref{future}, we briefly describe enhancements of the APS that
are being developed to expand the breadth of its
scientific capabilities.

\section{The Transformation of ALMA into a VLBI Station\protect\label{ALMArocks}}
ALMA is currently
the world's largest and most sensitive facility devoted to mm
and sub-mm astronomy (e.g., Wootten \& Thompson
2009). The ALMA array, located on the Chajnantor
Plateau in the Atacama desert of northern Chile, comprises 66
reconfigurable antennas, twelve of which are
7~m diameter parabolic dishes arrayed in a closely-spaced configuration and 54
of which are 12~m diameter parabolic dishes. 
The antennas, as well as the
ALMA baseline (BL) correlator, the ALMA Compact Array (ACA)
correlator, and the Central Local Oscillator are located at the Array Operations Site
(AOS) at an elevation of
approximately 5000~m, while science operations are conducted from an
Operations Support Facility (OSF) at 2900~m.

Antenna baselines within the ALMA array can
range from $\sim$9~m (within the ACA) to 16~km, depending on the array
configuration. However, relatively compact configurations (where most
antennas are on baselines of 
$\lsim$1~km) are the most desirable for phased array operations, since 
decorrelation of signals caused by variations in the atmosphere above
each of the
antennas is minimized. 

Fifty of ALMA's 12~m antennas comprise the so-called ``12~m Array'', which
in its standard operating mode is used in
conjunction with the 64-antenna BL correlator as a
connected element interferometer to
perform sensitive, high-resolution imaging observations. The sixteen
remaining antennas (twelve
7~m antennas and four 12~m antennas) comprise
the ALMA ``Compact Array'', which can be operated using a
separate, independent ACA correlator for the purpose of obtaining
total power measurements and 
enhanced imaging of extended sources or large-scale structures. Antennas from
the ACA may also be correlated by the
BL correlator along with antennas from the 12~m Array.

With an outlook toward future enhancements to ALMA's capabilities, the
ALMA BL correlator (Escoffier et al. 2007) was designed with
``hooks'' to allow future VLBI
observations (Baudry et al. 2012; see Section~\ref{PICS}). 
Importantly, the BL
correlator was designed to carry out these additional operations while at the
same time performing the usual correlation of signals from the
individual baselines. The APS was conceived to
take advantage of this framework and provide the additional hardware
and software components necessary to operate ALMA as a phased array
and record the resulting data streams in a standard VLBI format.


\section{Overview of the ALMA Phasing System (APS)\protect\label{overview}}
An antenna phasing system intended to operate at mm (or sub-mm)
wavelengths must be designed to accommodate the unique
challenges of observing at these wavelengths, including atmospheric
attenuation and
the rapid fluctuations in
the signal phases from astronomical sources caused by
tropospheric water vapor (e.g., Carilli, Carlstrom, \& Holdaway 1999). These 
factors are important even at a superb observing site such as ALMA's.
An additional consideration for the phasing system at ALMA
was that, at minimum, it would be
capable of meeting the needs and goals of mm VLBI experiments 
that require an
order-of-magnitude improvement in sensitivity for continuum observations
compared with existing capabilities.

The aforementioned goals led to a set of functional requirements for the APS that included:
(1) ability to phase up and sum up to 61 antenna signals;
(2) a capability for real-time assessment of phasing performance;
(3) ability to apply rapid phase adjustments based on data from water
vapor radiometers (WVRs);
(4) VLBI-quality frequency stability and timing;
(5) full bandwidth, dual polarization VLBI recording;
(6) circularly polarized data products (as required to facilitate VLBI data analysis); and (7)
$\ge$90~\% phasing efficiency. In practice, requirement (7) needed to
be relaxed for end-to-end operations (see Section~\ref{efficiency}).
On-sky testing and verification 
of the system to validate it against these various
formal requirements are discussed in Section~\ref{CSV}. As described in Section~\ref{future},
provisions were also made for additional features that could be implemented
at a later date.

Another important consideration was the need to design
and deploy the phasing system and ALMA VLBI capabilities
without impacting normal ALMA operations.
Furthermore, it was required that 
scientifically valid standard interferometry data
from the ALMA BL correlator be archived in parallel during APS operations, independent of
the VLBI recordings.

\begin{deluxetable*}{lll}
\tabletypesize{\footnotesize}
\tablewidth{0pc}
\tablenum{1}
\tablecaption{Characteristics of the ALMA Phasing System}
\tablehead{
\colhead{Feature} & \colhead{Specification} & \colhead{Remarks} }
\startdata
\tableline

Number of phased antennas$^{1,2}$ & $\le$61 & Number of phased antennas must be odd.\\

Equivalent collecting area & 4185~m$^{2}$ & Assuming 37 phased 12~m
dishes  and neglecting efficiency losses$^{2}$.\\

Frequencies of operation$^{3}$ & ALMA Bands 3 and 6 &
Extension to Bands 1 and 7 anticipated for ALMA Cycle 7.\\

Effective bandwidth per quadrant & 1.875~GHz & Four quadrants total; see Section~\ref{PICS}.\\

Aggregate bandwidth & 7.5~GHz & 1.875~GHz per baseband per
polarization\\

VLBI recording speed$^{4}$ & 64 Gbps & 4 Mark 6 units, each 16~Gbps; dual
linear polarizations recorded; 2 bit sampling\\

SEFD$^{5}$ & 65 Jy (at 3~mm)    & Assuming 37 phased 12~m
antennas.$^{2}$\\
           &  97 Jy (at 1.3~mm) & ...\\

Phasing efficiency$^{6}$ & $\sim$60~\% & Empirically derived value,
including all measured efficiency losses.\\

Flux density threshold for phasing$^{7}$ & $\gsim$0.5~Jy & For Bands 3 and 6
in ALMA Cycles~4 and 5.
\\

\enddata
\tablenotetext{1}{The maximum is set by the design of the ALMA BL
correlator (Section~\ref{PICS}) and the practice of designating at least two antennas
unphased comparison antennas (Section~\ref{approach}).}
\tablenotetext{2}{The anticipated number of 
antennas available for phasing  during ALMA Cycle~5 is $\sim$37.}
\tablenotetext{3}{The APS was commissioned for use in ALMA Bands~3 and
  6. However, it is capable of operation in any band provided that
  weather conditions are suitable.}
\tablenotetext{4}{The maximum recording rate available at ALMA's VLBI partner
sites during ALMA Cycle~4 was 2~Gbps for the GMVA and 32~Gbps for the
EHT. In Cycle~5, this rate is expected to increase to 64~Gbps for the
EHT sites.}
\tablenotetext{5}{The system equivalent flux density (SEFD) for Band~3
assumes an aperture efficiency of 0.71
  and a typical zenith system temperature of 70~K; for Band~6, the SEFD assumes an aperture
  efficiency  of 0.68, and a zenith system
  temperature of 100~K.}
\tablenotetext{6}{See Section~\ref{efficiency} for discussion.}
\tablenotetext{7}{Improvements in the handling of delays
  (Section~\ref{sec:issues}) will enable direct phase-up on
weaker sources, while a future passive phasing mode
(Section~\ref{future}) will enable VLBI on weaker
sources.}

\end{deluxetable*}

\begin{figure*}
\center{\fbox{%
\includegraphics[width=6.0in,angle=0]{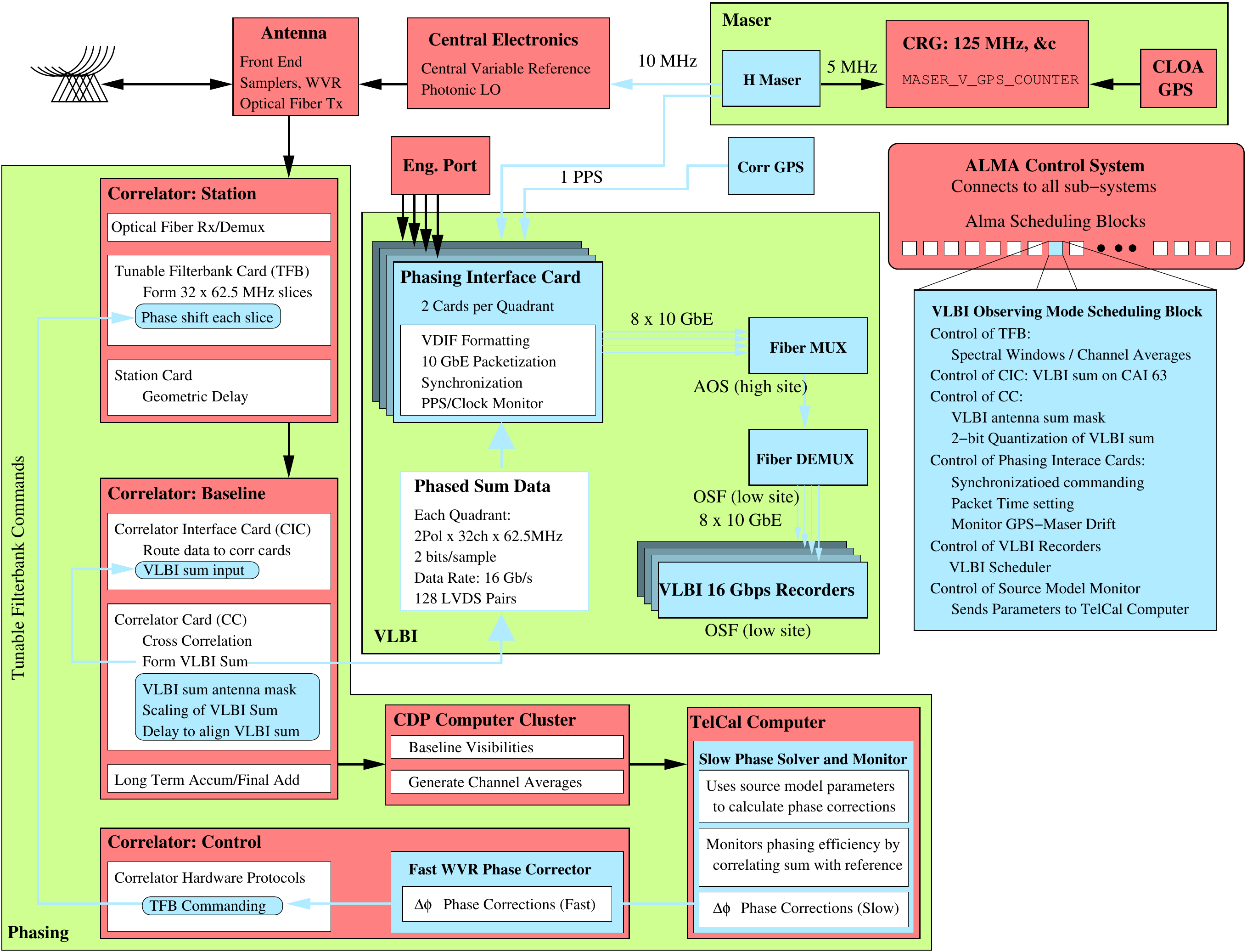}}}
\caption[ALMA Phasing Project System Diagram]{
Diagram of the APS, illustrating 
how existing ALMA systems were modified to support VLBI. Pre-existing 
objects are shown in pink and objects in blue are new components
added by the APP. The green
backgrounds group related components.}
\label{fig:app-hw}
\end{figure*}

The APS block diagram is shown in Figure~\ref{fig:app-hw}, and some of
the key characteristics of the integrated APS are summarized in
Table~1.
In Figure~\ref{fig:app-hw}, the green blocks group together elements into the three
main areas of hardware and software enhancements to ALMA: (i) frequency stability
(``Maser''); (ii) 
data formatting, transport, and recording (``VLBI''); (iii) the
beamforming of the array
(``Phasing'').
The phasing or ``beamforming'' component of the APS
analyzes the visibilities from the BL correlator to compute
phase adjustments to the individual antennas (Section~\ref{sec:SW}). This allows the formation
of coherent sum signals which are provided to specially designed phasing interface
cards (PICs) in the BL correlator (Section~\ref{PICS}).  At the same time,
these signals are sent through the BL correlator as if they
came from an ordinary ALMA antenna so that the performance of the
system may be monitored in real time.

\begin{figure*}
\center{\fbox{%
\includegraphics[width=6.0in,angle=0]{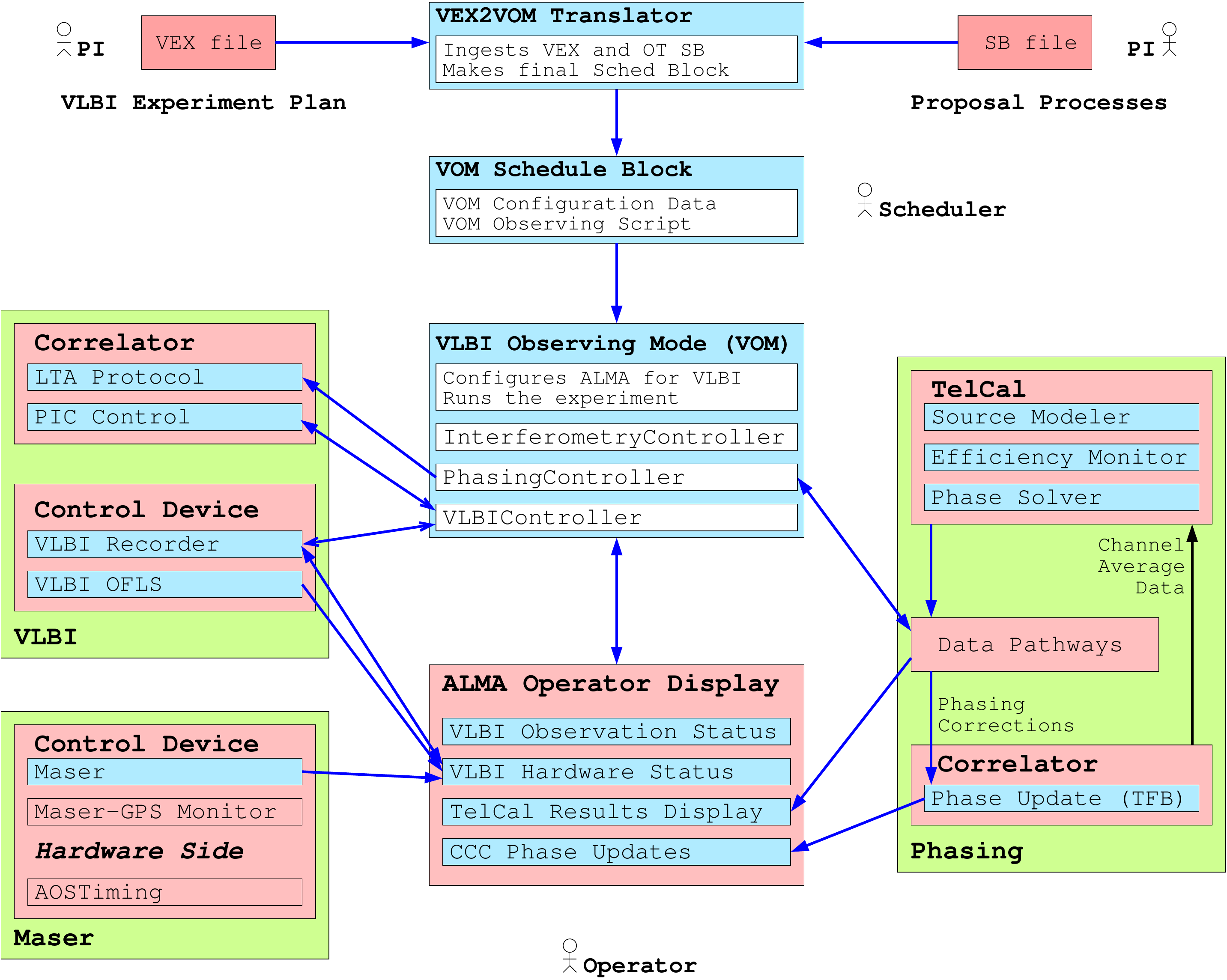}}}
\caption[ALMA Phasing Project Software Diagram]{
Diagram of the APS software, highlighting
the new components added to support VLBI.
Objects in pink previously existed; objects in blue are new and
were developed as part of the APP. The green
backgrounds group related components.}
\label{fig:app-sw}
\end{figure*}

In general, a VLBI DBE takes the
native analog receiver signals, downconverted to an intermediate
frequency (IF), digitizes them, and converts them to a standard
data format (in this case, VLBI data interchange 
format or ``VDIF'')\footnote{http://www.vlbi.org/vdif/} to allow
future playback at the VLBI correlators (Section~\ref{sec:korrel}). 
At ALMA the signals originate in the 
BL correlator at the AOS, and the  PICs serve as the VLBI backend,  translating the ALMA
digital signals into VDIF packets (Section~\ref{PICS}).  
These signals are brought down to the ALMA low
site (OSF) 
on an Optical Fiber Link System
(OFLS; Section~\ref{OFLS}) to the VLBI recorders (Section~\ref{mark6}).

All phasing operations in the APS are performed entirely in software.
The green fields in
Figure~\ref{fig:app-sw} delineate the three domains
of software activity required: (i) Phasing; (ii) VLBI Recording; and
(iii) Maser Control.
Modifications to the original ALMA software allow it to provide software ``devices''
to control and monitor all of the new VLBI hardware, as well as new
``controllers'' to manage the operations of these parts of the system.
One of these, the \texttt{InterferometryController}, is the same one
that runs standard (connected element) interferometry observations 
during normal (non-VLBI) ALMA operations.
The phasing system itself is managed by a
\texttt{PhasingController} 
(Section~\ref{approach}) that directs the BL correlator hardware
to make phasing adjustments based on calculations made in the Telescope Calibration
System (TelCal), which is the ALMA software component that provides
all online observatory calibration reductions. A \texttt{VLBIController} 
(Section~\ref{VLBIcontroller}) is also introduced to manage
the VLBI activities of the observation; it operates independently of the
\texttt{PhasingController}.  Figure~\ref{fig:app-sw}
also shows the software required to interface between the 
``Schedule Block'' normally used to execute ALMA observations, with 
the VLBI EXperiment (VEX) file paradigm that is a community 
standard for the execution of VLBI
observations (Section~\ref{vex2vom}).

A subsequent stage of VLBI operations is the 
VLBI correlation, which is performed after the
observations at specialized VLBI correlators using time-stamped data recorded at ALMA and
other observatories  (Section~\ref{sec:korrel}). The VLBI correlation 
is analogous to the role played by
the ALMA BL correlator during standard interferometry, but where data from the global VLBI
telescopes are processed instead of data from the connected ALMA elements. However, the time stamping
for VLBI requires  nanosecond (ns) 
precision, which is derived from a frequency-stabilized hydrogen maser
that was purchased for this purpose and adapted to meet all of ALMA's timing
needs (Section~\ref{maser}).  


\section{APS Hardware\protect\label{hardware}}
Implementation of beamforming and VLBI
capabilities at ALMA entailed installation of four new major hardware
components at ALMA: (1) a hydrogen maser to serve as a frequency
standard; (2) BL correlator modifications that included: (i) installation
of phasing interface cards (PICs) to serve as the VLBI backend and (ii)
installation of an additional card to serve as a 1 pulse-per-second (1
PPS) signal
distributor to the PICs;
(3) a bank of four
high-speed VLBI recorders; (4) an optical fiber link system (OFLS) to carry
the data stream from the ALMA correlator at the ALMA AOS to the recorders
at the ALMA OSF. 
We now describe each
of these components in more detail.

\begin{figure*}
\center{\fbox{%
\includegraphics[width=3.0in,angle=0]{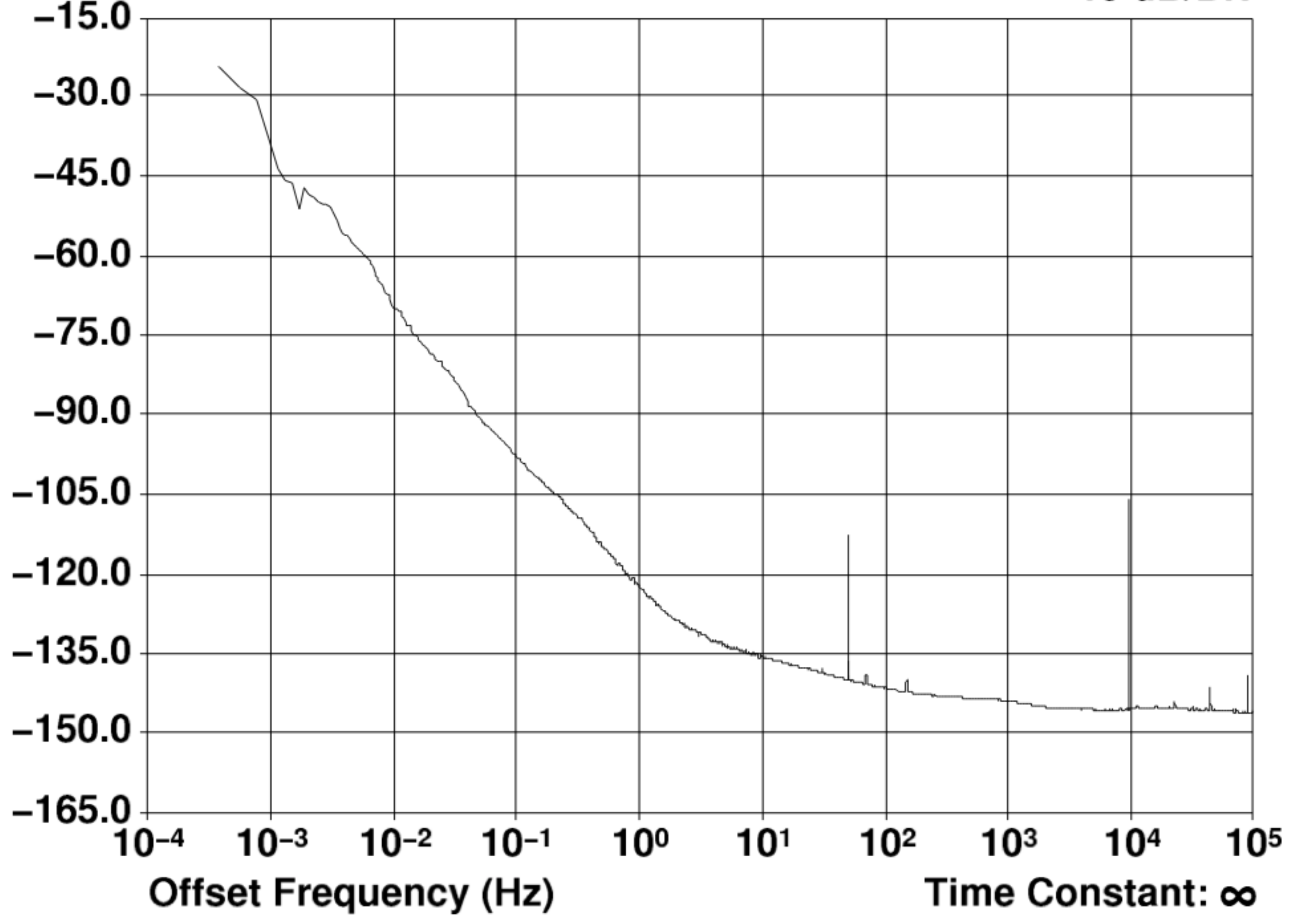}
\includegraphics[width=3.0in,angle=0]{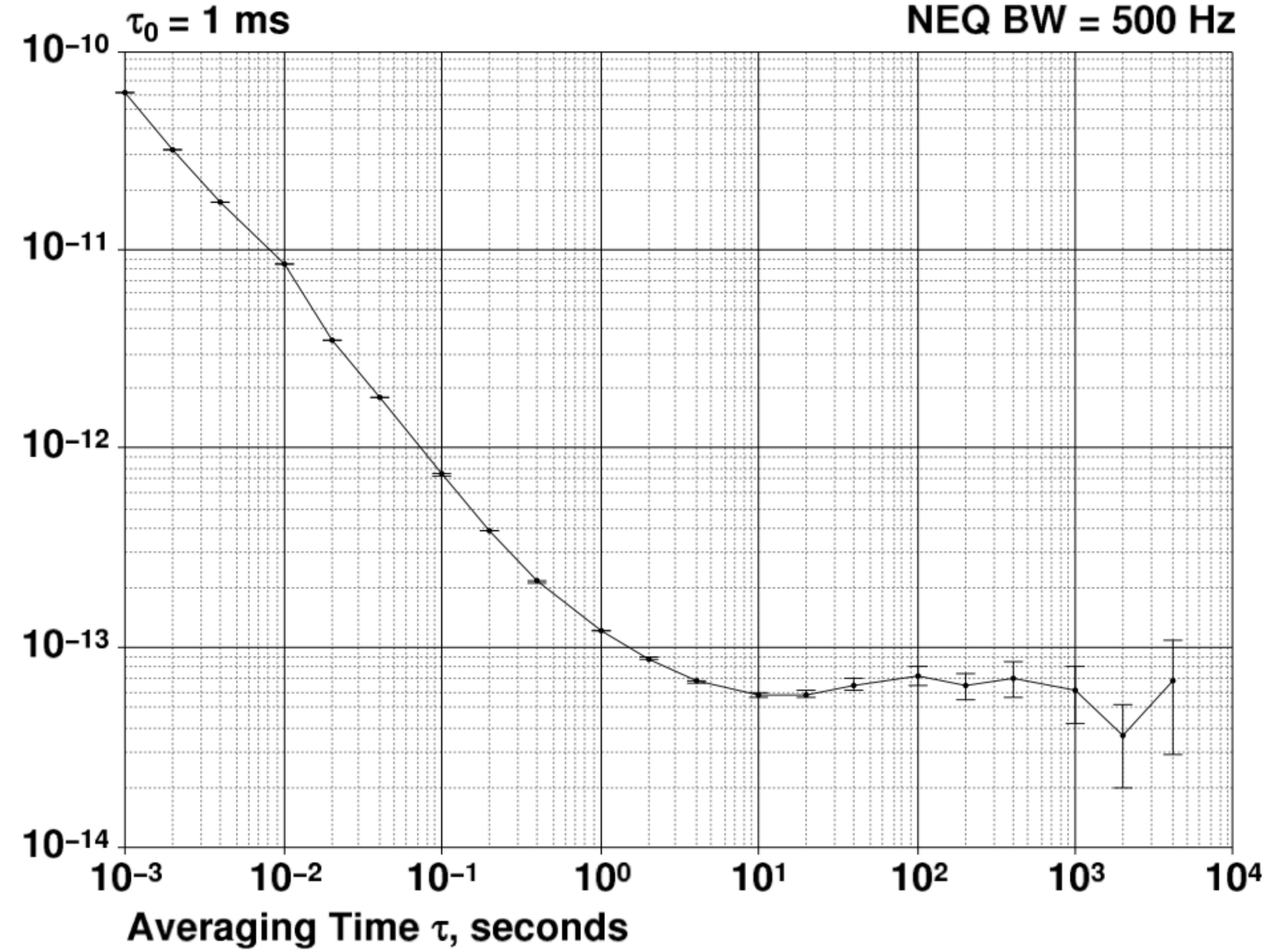}}}
\caption[Maser Phase Noise and Allan Variance Performance]{
 {\it Left:} Phase
noise  in units of dBc Hz$^{-1}$ as a function of offset frequency in
Hz based on a
comparison between the hydrogen maser installed at the ALMA AOS and a precision
quartz oscillator.  {\it Right:} Allan deviation between the
maser and the same quartz oscillator as a function of averaging
time in seconds. For a 1-second average, the Allan deviation for the
 ALMA maser alone is $1/\sqrt{2}$ times the value measured between the two
 devices, i.e., 
$8\times 10^{-14}$, which matches the required specification (see
Section~\ref{maser} for details).}
\label{fig:maser-allan}
\end{figure*}
\begin{figure}
\center{\fbox{%
\includegraphics[width=3.0in,angle=0]{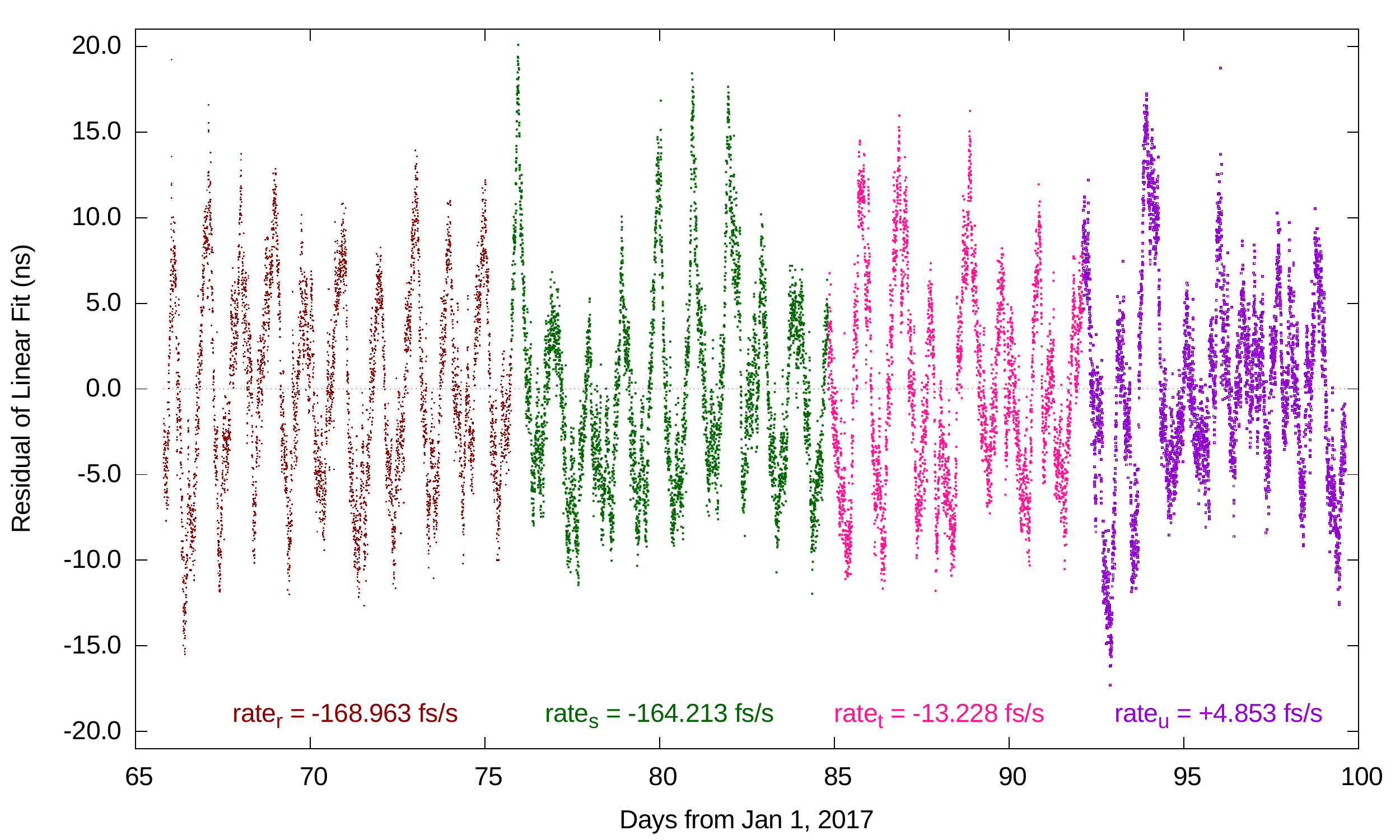}}}
\caption[Maser Timekeeping Relative to GPS]{
Timekeeping from the ALMA hydrogen maser relative to a GPS as a function
of time. Residuals to linear fits to the data, grouped in
approximately 10-day intervals (differentiated by color), 
are plotted for several weeks of data leading up to the ALMA
Cycle~4 VLBI campaign in 2017 April. These results show that there was
negligible drift in the maser over this interval.
}
\label{fig:maser-drift}
\end{figure}

\subsection{Maser\protect\label{maser}}
VLBI depends on the fundamental ability of radio receivers to
accurately preserve the phase 
of astronomical signals by referencing them to an independent frequency standard
at each of the participating VLBI stations.     
Prior to the APP, the ALMA array, when operating as a connected
element interferometer, depended on a common local oscillator
(LO) derived from a 5~MHz rubidium (Rb) clock. However, the Rb clock 
does not meet the strict stability
requirements for a VLBI frequency standard, namely
that the RMS phase fluctuations are kept well below one radian. 
This condition may be expressed as $\omega t \sigma_y(t) \ll 1$, where
$\omega$ is the LO
frequency in rad s$^{-1}$, $\sigma_y(t)$ is the Allan standard
deviation (equal to the square root of the
Allan variance), and $t$ is the integration
time in seconds (Rogers \&
Moran 1981).  

Because of their superb stability on time scales comparable to VLBI
integrations, hydrogen masers are used almost exclusively as
frequency references for VLBI.  At mm wavelengths,
tropospheric turbulence limits typical VLBI
integration times to $\sim10$~s.  Since modern hydrogen masers achieve
$\sigma_y(10~{\rm s})\approx1.5\times10^{-14}$, VLBI observations at 
wavelengths of $\sim$1~mm can be carried out with coherence losses of
$<5$~\% (Doeleman et al. 2011 and in preparation).  To fulfill this
need at ALMA, a
commercially available T4 Science iMaser was purchased and
installed in a seismic rack at the ALMA AOS as part of the APP. 
ALMA's Rb clock was also kept in place as a hot spare
for use in non-VLBI observations, but the hydrogen maser is now the
default frequency standard for all ALMA observing modes.
 
The original APS design called for a straightforward replacement of the 5~MHz
Rb reference with a 5~MHz output from the hydrogen maser. However, it was
found that to preserve the required VLBI stability and phase noise
specifications for the ALMA receiver LO signals, both the 5~MHz and 10~MHz outputs of
the hydrogen maser had to be utilized. This required a minor modification of
the ALMA Central Reference Generator, but the capability of switching back to
the Rb clock as a hot spare for non-VLBI observations was preserved.

The hydrogen maser at ALMA is routinely monitored in two
ways.   With the first method, the stability
of the maser's 10~MHz signal  is compared with that
of a high-precision quartz oscillator.  The crystal is not as
good as the maser over long periods, but it is of comparable
stability over $\sim$1-second intervals (where the maser signal is also
conditioned by a crystal).  Results of such a comparison 
are shown in Figure~\ref{fig:maser-allan}.
 
The left panel of Figure~\ref{fig:maser-allan} shows a phase noise
comparison between the ALMA  maser and an Oscilloquartz 8607
quartz oscillator. The phase noise specification for the
maser is $-$124~dBc Hz$^{-1}$ at 1~Hz from the 10~MHz reference. In a comparison of two
oscillators with similar noise characteristics, one expects to measure a value
that is 3~dBc Hz$^{-1}$ higher, and the phase noise plot confirms the
maser is operating at specification 
or better.  The right panel of Figure~\ref{fig:maser-allan} is the
measured Allan deviation, $\sigma_y(t)$, between the hydrogen
maser and the same quartz oscillator, both of which should have similar
stability over 1-second integrations. The measured value of the Allan
deviation of $1.15\times 10^{-13}$ at 1
second for two similar oscillators, implying that the Allan deviation for the
maser alone is $8\times 10^{-14}$ (i.e., $1/\sqrt{2}$ times the value
measured for the two oscillators), 
which matches the required specification. 

A second type of maser health check involves routinely monitoring the offset between
the 1~PPS signal
generated by the maser from its 10~MHz output with another 1~PPS signal generated by a global
positioning system (GPS) receiver.   GPS stability on 1-second
timescales is poor due to variations 
induced by the ionosphere  and will vary by a few tens of nanoseconds on a diurnal cycle
due to ionospheric total electron content (TEC) changes. However, on longer
timescales ($> 10^4$~s), GPS is providing an average time rate synchronized with the United States
Naval Observatory timekeeping
for Coordinated Universal Time (UTC) and its
precision will exceed that of a single hydrogen maser.  If the maser is
operating properly, a plot showing a comparison of the two signals is
expected to show a  
linear drift at a rate characteristic of the maser, with some arbitrary offset
due to the initialization of the first maser 1~PPS pulse.  A plot of the
residuals to such a linear fit for the ALMA maser is shown 
in Figure~\ref{fig:maser-drift} for
the period leading up to the 2017 April VLBI campaign at ALMA.
It shows several weeks of data, with fits 
on $\approx$10-day intervals (designated in different colors).  
Prior to the campaign, the maser was
``tuned'' by adjusting the digital synthesizer which generates the
10~MHz oscillation for zero drift relative to UTC.  Several fits
over the final two-week period suggest that the average drift was essentially
zero, with a measurement uncertainty of fs s$^{-1}$.

\subsection{Baseline (BL) Correlator Modifications\protect\label{PICS}}
The 64-antenna ALMA BL correlator processes four 2~GHz frequency 
chunks of dual-polarization spectrum from up to 64 antennas, one chunk per 
quadrant. Each data stream requires a
phasing interface card (PIC) to serve as a VLBI backend and to transfer 
the ALMA signals to the VLBI recorders. In total, eight
PICs are required (one per polarization in
each of the four 2~GHz correlator quadrants). 

\begin{figure*}
\center{\fbox{%
\includegraphics[width=6.0in,angle=0]{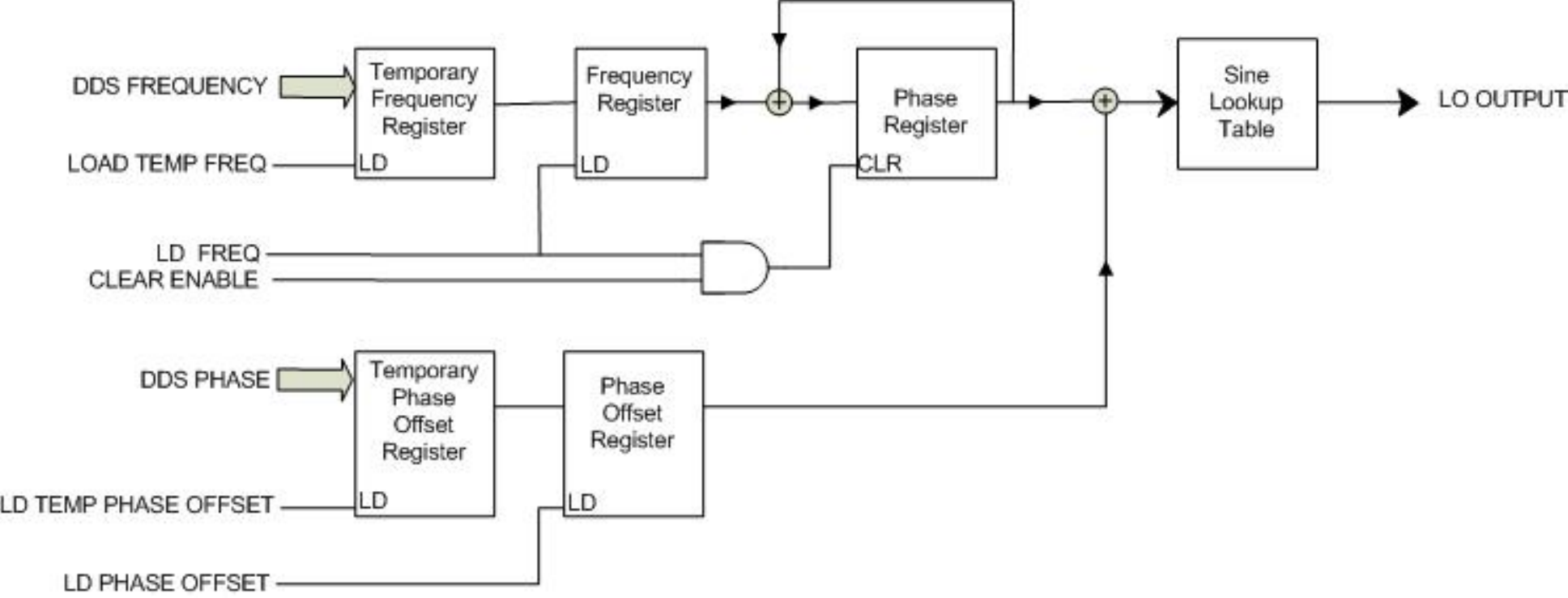}}}
\caption{TFB Phase Adjustment Logic. 
In normal operation of each TFB, the logic in the upper half of this
figure is active:  the desired frequency is loaded and the contents
of the phase register are cleared at the start of the observation.
Thereafter the LO output is driven from a sine lookup table.  When
the APS is active, an additional 16-bit phase offset register is
loaded with a value for the appropriate fraction of a turn of phase
every correlator subscan in order to correct the phase of this TFB.
This phase offset may be cleared (or not) at the start of a scan.
The capability of leaving the phase offset register contents
 unchanged will enable a future use of a ``passive'' phasing mode 
(Section~\ref{future}).
}
\label{fig:tfb-register}
\end{figure*}

\begin{figure*}
\scalebox{0.7}{\rotatebox{0}{\includegraphics{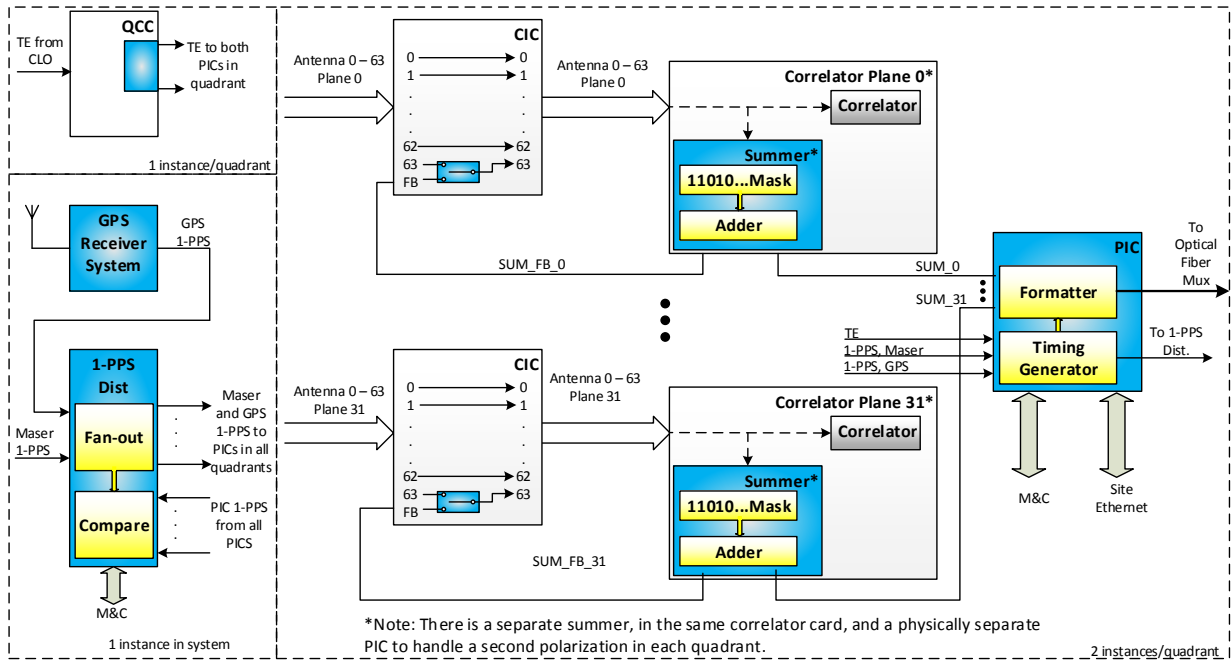}}}
\vspace{-5.5cm}
\caption{Block diagram of the ALMA BL
correlator modifications required for the APS. Changes are shown with blue boxes.  Timing
is managed with Timing Events (TEs) every 48~ms.
The Quadrant Control Cards (QCCs, upper left box) were
modified to send these signals to the eight Phasing Interface Cards
(PICs), as depicted along the right.  To verify timing, 
an independent GPS receiver and distribution system for a
1~PPS signal (lower left)
were added.  In addition, a 1~PPS signal from the
maser (Section~\ref{maser}) is distributed to each PIC.  The Correlator Interface
Cards (CICs; two are shown on the left side of the main box) received
a logic switch in the gateware that switches Correlator Antenna
Input (CAI) number 63 between a single ALMA antenna (normal operations)
 and the output of the summer logic in the Correlator Plane
assemblies.  The summer has a mask to determine which antennas
contribute to the sum, and it adds the appropriate 
2-bit signals and rescales them back to two bits for input
at CAI~63.   The sums are performed on
  a per tunable filter band (TFB) channel basis, with 32 TFBs per
  quadrant in each polarization
(SUM\_FB\_0 ... SUM\_FB\_31).
 }
\label{fig:rich}
\end{figure*}

The ALMA BL correlator has an ``FXF'' architecture (Escoffier et al. 2007).  
The 2~GHz input band is initially sliced into thirty-two 62.5~MHz
pieces 
(``F'') by sets of tunable digital filter bank  (TFB) cards, whose
design is described in Camino et al. (2008).  Each 62.5~MHz ``TFB channel'' has an
approximately flat response determined by the digital filter design
implemented in the field programmable gate arrays (FPGAs) of each TFB
card. These channels are overlapped slightly in frequency (15/16th of
each channel), yielding an effective 
bandwidth is 1.875~GHz per quadrant.
The TFB channels are processed in separate correlator ``planes''.  
The TFBs are followed by a hardware correlator (``X'') 
and software Fourier transform (``F'').  
The correlator was designed with ``hooks'' for VLBI (Baudry et al. 2012), 
including spare card slots and rack space; extra power capability;
a digital, high-resolution adjustment capability for the
phases of the LOs in the TFBs
(to 1/4096 of a 360$^{\circ}$ turn; 
see Figure~\ref{fig:tfb-register}); and FPGAs dedicated to computing a sum of a
selectable set of antennas.

The BL Correlator has two modes of operation with regard to
how the antenna signals are passed through the TFBs: time division
mode (``TDM'') and frequency division mode (``FDM''). TDM is the
default for continuum observing with ALMA during standard
interferometry observations; in this mode, the TFB cards are bypassed
and the correlator behaves like
a pure XF rather than an FXF system (Baudry et al. 2012).
However, only FDM allows for phase adjustment of the TFBs and the option to
easily capture channelized data in a  VLBI data set. 

In practice, the spectral specification allows one to define some
number of channel averages (spectral sub-regions), which correspond
to sets of TFBs, for the purpose of making phase adjustments. Because
the employed FDM setup generates more ALMA standard interferometry data  than can
readily be archived in real time, some degree of spectrum averaging (typically a
factor of eight)
is used to reduce the volume of archived data. 

The final design of the correlator 
modifications required for the APS is shown in Figure~\ref{fig:rich}.  
Parts added for VLBI are shaded blue, with additional detail shown in yellow.
Because timing precision and stability are critical to successful VLBI
operations,
a 1~PPS distribution system (card and cables) was installed
to bring signals from the hydrogen maser (Section~\ref{maser}),
as well as from a GPS unit, to the PICs 
so that the timing of the data in the packets could be continuously recorded
against both of these signals (left side of Figure~\ref{fig:rich}). 
The 1~PPS signal from the maser should be completely
invariant in the data, since the entire ALMA system is operated from the
maser 10~MHz output.  On the other hand, as noted above, the 1-PPS signal 
from the GPS unit typically varies by tens
of nanoseconds due to diurnal ionospheric TEC changes. 

One complicating  factor in the adaptation of the ALMA BL correlator for
VLBI is the fact that ALMA uses a 
48~ms Timing Event (TE) instead of the VLBI standard 1~PPS signal as 
a timing epoch standard. For this reason, 
several redundant measures were designed to insure that the timing 
is correct.  The maser 1~PPS signal, a GPS 1~PPS signal, and the site TE are 
connected to the Timing Generators in each of the eight PICs. 
By design, the TE is coincident with the 1~PPS signal at 
seconds 0, 6,...54 and is carefully distributed so that 
its rising edge is within the same 125~MHz clock cycle everywhere in
the correlator.  To synchronize the PIC 1~PPS signal, the real-time control 
computer sends a ``synchronize'' command to each PIC's Timing Generator 
just prior to the TE at second 0.  The 1~PPS Distributor has the
ability to cross-check all of the PIC-generated 1~PPS signals against 
both the GPS and maser 1~PPS signals. 

The ALMA BL correlator contains 64 Correlator Antenna Inputs (CAIs),
numbered 0 through 63. 
To support VLBI, the firmware on the Correlator Interface Cards (CICs) was
modified by insertion of a logic switch that toggles
the input at CAI \#63 from a single ALMA antenna to 
a phased sum of the array antennas (see Figure~\ref{fig:rich}). 
This allows the phased sum to be correlated 
against any individual array antenna, providing a means of validating that the 
summation is correct.

The sum signal itself is produced using the summer logic in the Correlator Plane
assemblies (Figure~\ref{fig:rich}). 
The analog sum logic (gateware) in the dedicated FPGAs on
each Correlator Card was modified to
properly scale the outputted digitized samples and produce 2-bit signals for the PICs and
CIC feedback. The sum is formed on a per TFB channel basis 
using a mask-selectable set of $N_{\rm A}$
antennas, where the mask is user-defined and relayed to the system via the
correlator control system.

The mask is required for multiple reasons.
First, in order to faithfully represent the sum, $N_{\rm A}$ must be odd. 
The reason is that in the available hardware circuit,
only two bits per antenna are available. This allows representation  of
the  signal voltage (part of a Gaussian distribution) as one of
four possible values (two positive and two negative states).
Consequently, with an even number
of antennas, the resulting sum can be exactly zero, which in
turn cannot be represented by any of the four available states.
The simplest solution is to impose the
requirement that the number of antennas be odd. Because a ``zero'' state
is in fact the most probable one for a sum of an even number of antennas,
the SNR loss incurred by omitting one antenna from the phased sum is always smaller than
would result from having signals end up in non-representable zero states.
A second reason for the mask is that it is useful to hold some antennas apart from the sum
signal; one or more of these may be used as ``comparison antennas'' to
estimate the efficiency of the phasing system (Section~\ref{efficiency}). Finally, 
depending on the configuration of ALMA, antennas that
are on long baselines ($>$1~km)  may
be excluded to help insure good phase stability across the selected
array.

In addition to the Timing Generator discussed above, the PICs also include
a formatter.  The formatter takes 2-bit data streams from 32 of 
the Correlator Cards in each quadrant and for each polarization, integrates them into a
VDIF-compliant data stream, and routes them to a 10~GbE interface. The
VDIF data stream is packaged in User Datagram Protocol (UDP) packets onto a 10~Gbps optical
fiber which is mated to the PIC using an SFP+ network adapter.
The PIC formatter supports bandwidths from 62.5~MHz to 2~GHz in binary multiples of 
62.5~MHz and can also produce several types of test patterns for 
troubleshooting. 

\subsection{Mark 6 VLBI Recorders\protect\label{mark6}}
The Mark~6 VLBI recorder is the current offering in a long line of VLBI recording systems
developed by the Massachusetts Institute of Technology (MIT) 
Haystack Observatory. It is commercially manufactured by Conduant,
Inc. The Mark~6 is based on commercial off-the-shelf
technology and open-source software. Data are recorded
onto ``modules'' comprising banks of eight commercial-grade
3.5-inch hard disks inserted into a Mark~6 chassis. 
Typically the disks are ``conditioned'' prior to use to eliminate bad
sectors and to verify their performance.

\begin{figure}
\center{\fbox{%
\includegraphics[width=3.0in,angle=0]{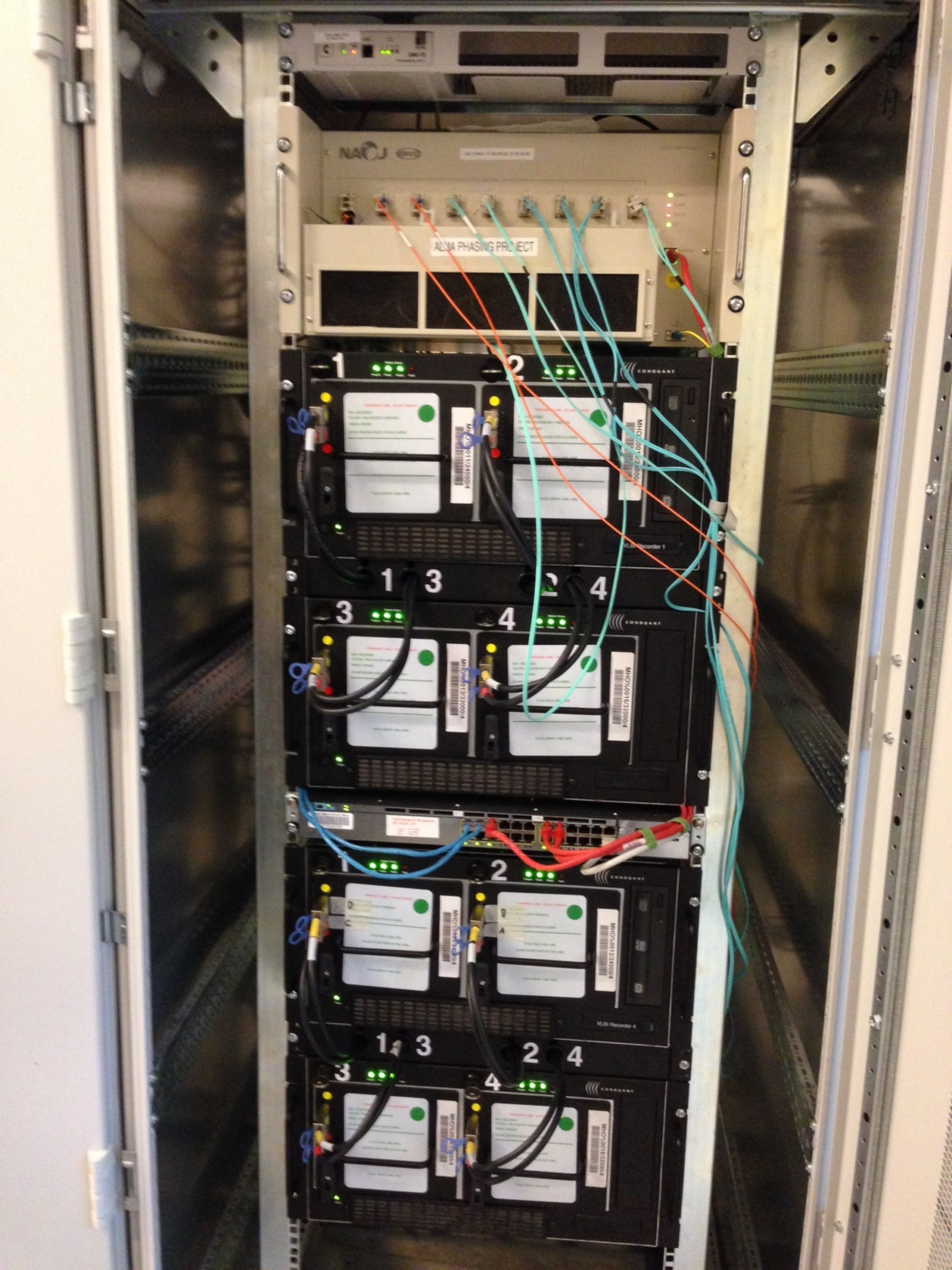}}}
\caption{Recorders 1 and 4 installed in the designated ALMA OSF computer room rack.
The OFLS unit is mounted at the top.   Between the recorders is the network
switch which connects the recorders to the AOS network and also provides a private
network between the recorders.}
\label{fig:racked}
\end{figure}

Mark~6 recording was initially demonstrated in 2012
(Whitney et al. 2013) and first adopted for astronomical VLBI use
at ALMA. Figure~\ref{fig:racked} shows two of the Mark~6 recorders
shortly after installation at the ALMA OSF.
In the ALMA configuration, each recorder accepts two VDIF packet
streams (one per polarization) on two optical network
cables and writes data onto four disk modules at an aggregate
rate of 16~Gbps.  When all four ALMA basebands are in use,
the total recording rate is 64~Gbps (2~GHz of bandwidth from each of four basebands, 
Nyquist sampled, 2 bits per sample, with dual polarizations).

In normal operations, there is some manual effort required
to mount and unmount the disk modules and verify that these
have survived the stress of international shipping. Thereafter
the operation of the recorders is completely automatic at
the ALMA Control system via the observing
script and the \texttt{VLBIController} software.

The Mark~6 system development was concurrent with the APS development,
and there were  a number of issues to consider
in adapting the recorder for use at ALMA.  We needed to ensure that the
capacity of the modules was sufficient to eliminate the need for
module swaps during overnight 
VLBI observing shifts. Because the recorders were sited at the ALMA OSF
(elevation 2900~m) in a climate-controlled computer room, some environmental issues
relevant to other VLBI sites were not an issue.  Nevertheless, we
undertook extensive
testing to verify that ordinary hard drives (rated for
10,000~ft) were indeed usable at ALMA.

In contrast to previous hard-disk VLBI recorders which used proprietary
disk formats, we opted for an open, simple, non-RAID, ``scatter-gather''
plan for distributing data in $\approx$10~MB blocks across multiple disks.
In practice, at ALMA the data from each PIC are thus spread over 16
disks so that a useful recording is still possible even in the event
of network issues or disk failures.
Finally, since the Mark~6 recorders are conventional Linux systems, it
is possible to run DiFX correlation software (see
Section~\ref{sec:korrel}) on the mini-cluster comprising the
four recorders in order to perform local VLBI correlations of data.  This was
valuable for early testing, although in practice, the ability to routinely
perform fringe tests with remote sites in near real time is precluded
by the limited network bandwidth between ALMA and the other sites.

\subsection{Optical Fiber Link System (OFLS)\protect\label{OFLS}}
An optical fiber link system (OFLS) is required to transfer the antenna sum data 
generated by the APS from the ALMA AOS to the ALMA OSF where the VLBI recorders are
located. The OFLS was designed by the National Astronomical Observatory of
Japan and Elecs Industry Co. LTD. and
fabricated by 
Elecs Industry.
The system multiplexes the eight digital network streams generated by
the PICs in the BL correlator (Section~\ref{PICS}) onto a single ALMA fiber (out of a 
bundle of 48 fibers) for transmission. The measured signal loss ranges from
$\sim 6-7$~dB over the $\sim$30~km cable path.
At the two ends, identical units multiplex the signals onto the single 
fiber with dense wavelength division multiplexing.

\begin{figure*}
\centering
\includegraphics[width=10cm,angle=-90]{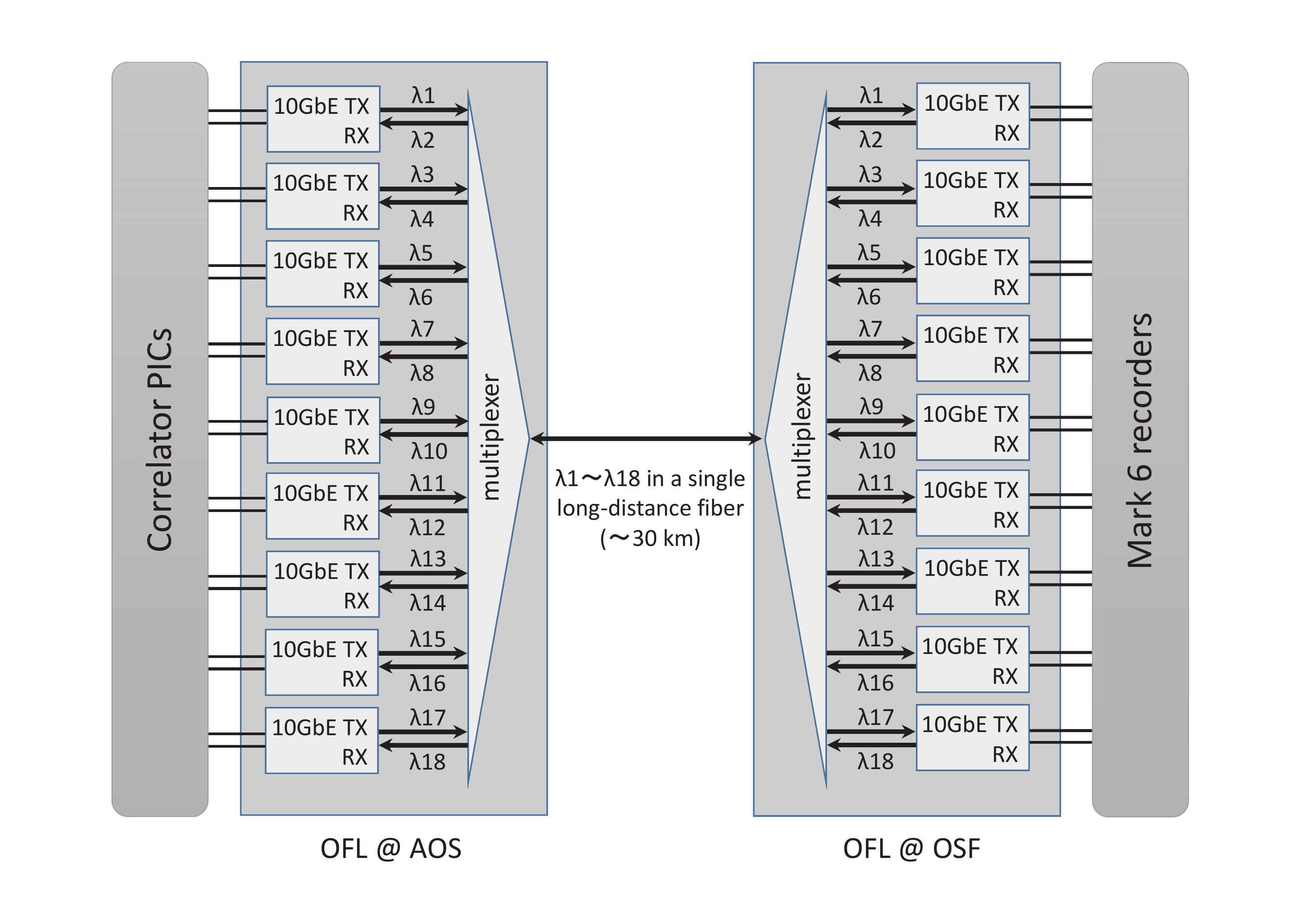}
\vspace{-0.5cm}
\caption{A schematic diagram of the optical fiber link system (OFLS),
  which connects the data stream generated by the BL
  correlator at the ALMA AOS with the Mark~6 VLBI recorders at ALMA OSF.}
\label{fig:ofls}
\end{figure*}

Each of the eight channels of the OFLS has a maximum data rate of 10~Gbps,
allowing it to accommodate the data
rate generated by each of the eight PICs (8~Gbps, plus a small overhead for the
packet header of $<$1~\%).
There is also a ninth channel available as a built-in spare, so 
in total there are 18 different data streams: nine downlink streams and
nine uplink streams. The OFLS design is bi-directional since it is
built based on 10GbE technologies, although the uplinks are not used during observations.

Figure ~\ref{fig:ofls} shows a schematic diagram of the OFLS.
The data transfer through the single long-distance fiber is done with 
a 10GBASE-ZR optical module operating at a wavelength of 1550~nm.
The optical channels designated $\lambda_1$ to $\lambda_{18}$ are assigned to
the International Telecommunications Union  
(ITU) grids  of 21, 23, 25, ..., 55, corresponding to
frequencies of 192.1, 192.3, 192.5,...195.5~THz (i.e., 100~GHz spacing).
In the OFLS the optical data are converted between 10GBASE-ZR and 
10GBASE-SR so that the OFLS can interface the correlator PICs and
Mark~6 recorders with 10GBASE-SR optical modules. 

Software connection to the OFLS and network control can be made with the
VLBI Standard Interface for Software (VSI-S) protocol on
a physical 10BASE-T/100BASE-T ethernet.
For monitoring its status, the OFLS provides information such as 
power status, the link status of each channel, and temperature status.
It has no packet monitoring capability, and thus it is a passive 
instrument in the VLBI phasing system.
The OFLS was built with redundancy in terms of the number of cooling fans and DC power supplies
to avoid failures in these parts, which tend to have relatively short
mean times between failures
(between a few to several years).


\section{APS Software\protect\label{sec:SW}}
\subsection{The VLBI Controller and VLBI Observing Mode (VOM)\protect\label{VLBIcontroller}}
A VLBI Observing Mode (VOM) at ALMA is required to support the
VLBI back end; this is handled by the
\texttt{VLBIController} as indicated in Figure~\ref{fig:app-sw}.
By design, the VOM performs only limited functions: 
the PICs need to be programmed and the recorders need
to be told what to record.  For the PICs, programming consists
of providing them with the VDIF header, which includes the number
of channels to record and, critically, the timestamp for the
data.  As discussed in Section~\ref{PICS}, 
the ALMA system is time-managed around TEs which occur every 48~ms.  Every six seconds, one
of these is coincident with an integral second, so one
of these coincidences is chosen for the moment of programming:
a VDIF header is downloaded prior to the TE and at the TE, the
PIC starts its packaging and broadcasting of data.  This stream
is generated continuously until the correlator is reset or
the PICs are turned off.

The data are only useful during
the parts of the observing
sequences when the array is phased-up, as described below. Therefore
at the start of the observation, the recorders are provided a schedule
from the VEX file appropriate for the current set of
observations and the recorders then operate autonomously.  Since
the \texttt{VLBIController} also has access to the schedule, it can ``check in''
with the recorders and report the result of a scan check that
verifies that the data have been recorded properly.

\subsection{The Phase Solver\protect\label{solver}}
To exploit the full sensitivity of ALMA for VLBI experiments, it is necessary to operate the ALMA
array as
a virtual single dish rather than an interferometer. This is
accomplished by coherently summing the voltages $V$
received at each of the $N_{\rm A}$ individual dishes:

\begin{equation}
V_{\rm sum} = (V_{1} + V_{2} + ... + V_{N_{\rm A}}).
\end{equation}

\noindent  Here the voltage from a given antenna, $i$, is the
combination of a signal voltage $s_{i}$ and a noise voltage
$\epsilon_{i}$, i.e., $V_{i}=(s_{i} + \epsilon_{i})$. For the
  APS, $N_{\rm A}$ must be an odd number because of the 2-bit quantization
  of the signal (see Section~\ref{PICS}). The power
from the combined signals of all possible baselines $ij$
may be expressed as:

\begin{equation}
\langle V^{2}_{\rm sum}\rangle = \sum_{ij}[\langle s_{i}s_{j}\rangle +
\langle s_{i}\epsilon_{j}\rangle + \langle s_{j}\epsilon_{i}\rangle + \langle\epsilon_{i}\epsilon_{j}\rangle]
\end{equation}

\noindent (e.g., Thompson, Moran, \& Swenson 2017; hereafter 
TMSIII). In the case where the individual antenna
phases are random, their signals will be uncorrelated with those of
other antennas. Thus the only non-zero term in Eq.~2 will be $\langle
s_{i}s_{j}\rangle$ for the case $i$=$j$, and the combined array becomes
effectively equivalent in sensitivity to that of only a single element
$i$. In contrast, if the signals are optimally phased before summing
them, a coherent addition of the signals results in $N_{\rm A}$ times the
unphased array sensitivity (e.g., Dewey 1994), minus efficiency losses
caused by atmospheric effects and quantization of the signals before summation
(see Section~\ref{CSV}). It is for this purpose of performing this
optimized sum that
specialized software needed to be developed to enable phase-up the ALMA
array. The use of these phase solutions is discussed in the next subsection.
 
\subsubsection{Approach\protect\label{approach}}
\begin{figure}
\center{\fbox{%
\includegraphics[width=8.5cm,angle=0]{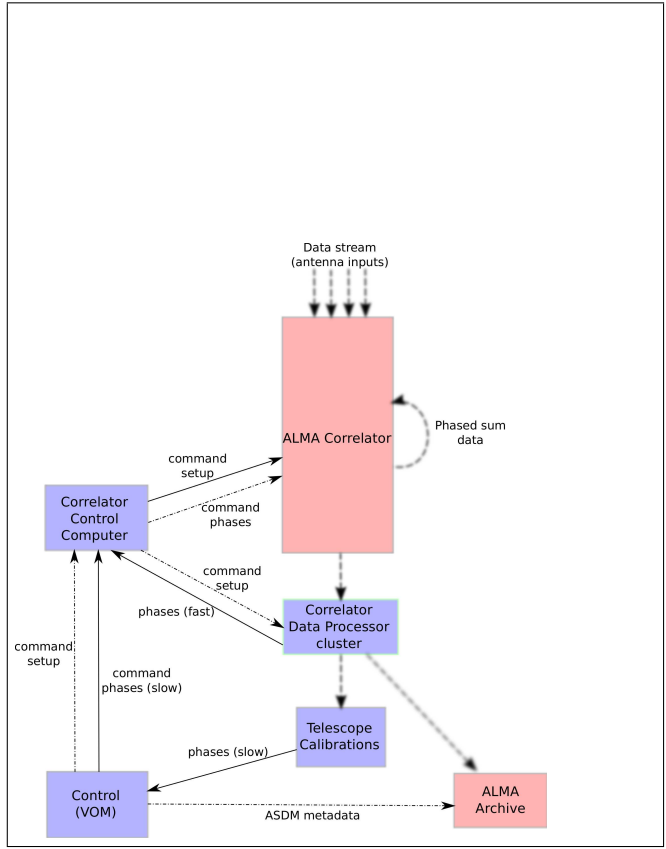}}}
\caption[Simplified View of Phasing Loops]{
Simplified view of the ALMA phasing loops, with
software elements indicated in blue. Bold, segmented
arrows depict the stream of binary data from the antennas
to TelCal and the ALMA Archive; other arrows represent
the transfer of control data and commands.  Adapted from Mora et
al. (2014). }
\label{fig:phasing-loop}
\end{figure}

The phasing software for the APS was custom-designed for the existing ALMA BL
correlator hardware and software. Figure~\ref{fig:phasing-loop}
provides an overview of the flow of
information among the components of the APS software. 

In general, the BL correlator
generates interferometric
visibility data on a ``subscan'' basis  (where a subscan has a duration of order
seconds); these are normally
gathered into ``scans'' (of order minutes) for publication to the ALMA
Archive, where the raw visibility data are deposited and stored 
for eventual scientific analysis.
Certain calibrations (performed in TelCal) are also routinely applied on
this data stream as requested by the observation software.
This is the purview of the
\texttt{PhasingController} shown in Figure~\ref{fig:app-sw}.

 The BL
correlator hardware is set up by the correlator control computer (CCC), and once
programmed, it continuously correlates the input signals from all 
antennas in the observation. Antennas are managed as an ``Array'' by the
control system (e.g., for the purpose of commanding pointing at
targets), and the signals from the defined
Array are correlated to provide cross-correlations between all antenna pairs
(baselines). This computation also contains the antenna phases
relative to one arbitrary reference antenna in the Array (see Section~\ref{sec:slow-loop}), as
needed to coherently sum the antennas on the next trip through the BL correlator.

\begin{figure*}
\center{\fbox{%
\includegraphics[width=13.5cm,angle=0]{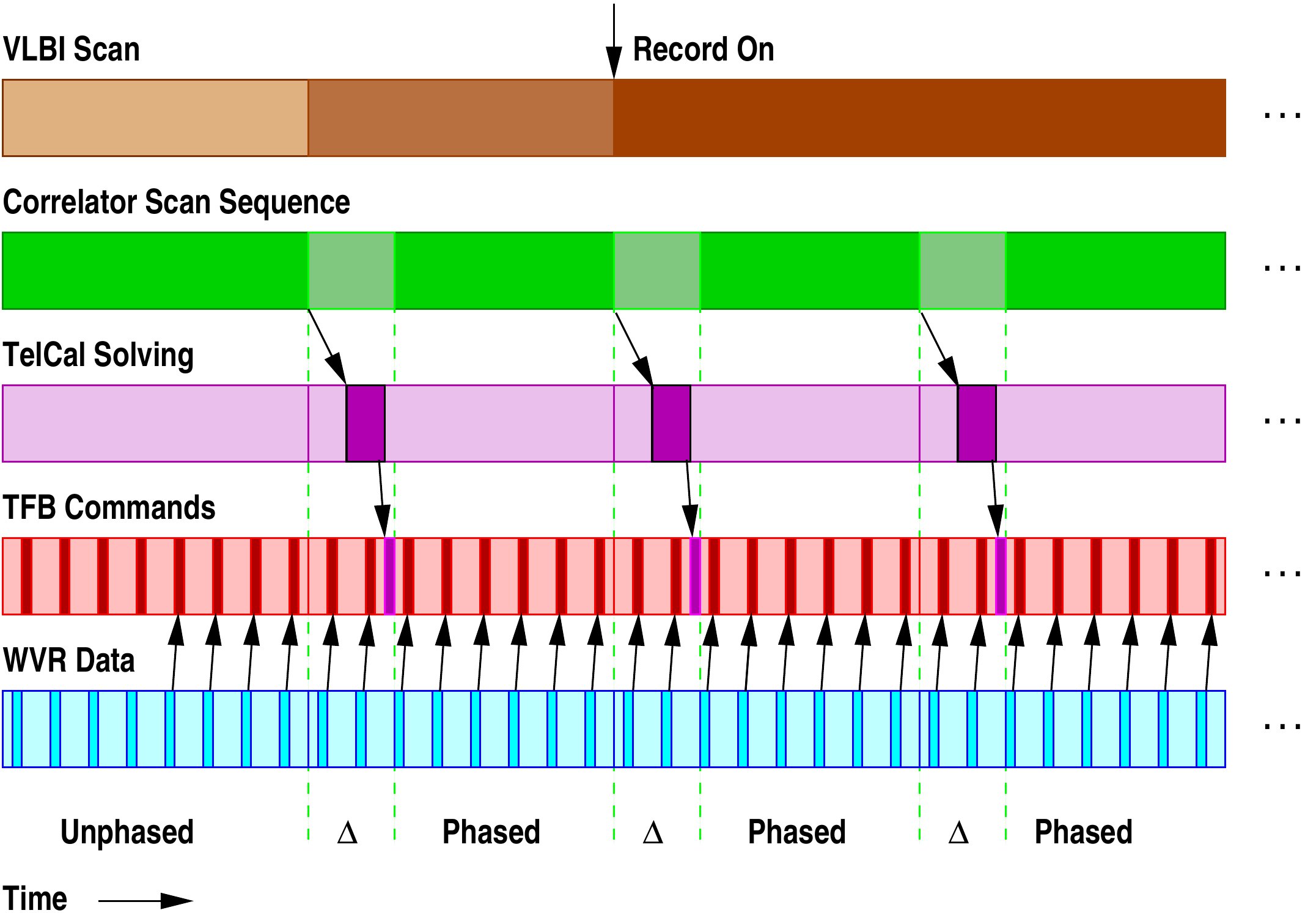}}}
\caption[Scans in the Phasing System]{
Schematic illustrating the relationship between different types of
scans in the APS.
A ``VLBI scan'' is partitioned into ``subscans'' for correlation and
for processing in TelCal (the ``slow'' phasing loop).  Optionally, 
WVR adjustments are made
in the CCC on a shorter timescale, of order 1 second (the ``fast''
phasing corrections; see Section~\ref{sec:SW} for discussion). The green vertical lines
designate the portion of the ALMA scan sequence during which the slow
phasing corrections are computed and published.
Darker shaded regions within each row depict the relative duration of the various commanding
operations. Adapted from Mora et
al. (2014).}
\label{fig:phasing-scan}
\end{figure*}

Figure~\ref{fig:phasing-scan} illustrates the temporal relationship
between the 5 timelines that occur during a VLBI scan. Information is
processed on two timescales or ``loops'', which may be deployed
individually or in parallel. The arrows in the figure refer to transfers
of information in these two loops.  The ``slow'' loop
(Section~\ref{sec:slow-loop}) processes the
visibility data and applies corrections on the order of every 10 seconds
based on phasing results from the previous scan.  
It is a closed loop between the BL correlator, where
measurements are made and applied, and TelCal, where corrections
are calculated.  The ``fast loop''
(Section~\ref{sec:fast-loop}) operates on timescales of
order one second. It applies corrections based on data from the WVRs associated
with each 12~m antenna, which
measure additional delays in the signal path due to water vapor above each telescope.

Of the 64 CAIs, numbered 0 to 63,
one (\#63) is reassigned to hold the phased sum (see Section~\ref{PICS}). 
Since the sum antenna is not a real antenna, it cannot be pointed,
so it is not part of the Array at the control level.
However, once the sum is calculated by the CIC, the
BL correlator can cross-correlate it as if it were
a real antenna.

The BL correlator spectral processing software makes one delay adjustment for the
192~ns required to perform the summation and route it back to the
correlation inputs.  The correlations are then dumped from the correlator
boards into the long-term accumulator (LTA) cards and ultimately read by the nodes
of the Correlator Data Processing (CDP) cluster, which is comprised
of four slaves per quadrant and a master, which commands the slaves.
This is where the final spectral processing is done prior to delivery
of the data to the Archive.  The full set of correlations
is not present in any one computer until the Archive stage.
The four slaves are assigned correlation work based on the CAI numbers
of the antennas at the ends of each baseline.  One slave
handles baselines where both antennas have CAI$<$32, another
handles baselines where both antennas have CAI$>$31, and the
other two handle the baselines where one CAI$<$32 and the
other has CAI$>$31.

Since there is a significant processing load on all of these machines,
as well as the networks connecting them,  the phasing calculations are
done instead in TelCal. For each subscan, TelCal performs
the necessary calculations to phase the array and inform the VOM, which then relays the
commands needed for application of the phasing corrections by the CCC.  
The WVR data (supplied
through the CAIs from the antennas) are provided to the CDP slaves
so that online spectral processing can optionally include WVR
(fast-loop) corrections.
If selected, these latter corrections are applied at a cadence
of about once per second and are directly forwarded to the CCC.
All of this is managed by the \texttt{PhasingController}
through commands to the CCC, TelCal, or the CDP master.  A complete
record of these activities is provided in an ALMA Science Data Model
(ASDM) table that is
stored in the Archive along with all the other
metadata from the observation.
The  \texttt{PhasingController} ensures that the
number of phased antennas is odd, and the
observing script (see Section~\ref{vex2vom}) generally specifies that
at least two antennas are omitted from the phased array, making them available as
comparison antennas to evaluate the performance of the phasing system
(Section~\ref{efficiency})
and facilitate calibration of the VLBI data. Thus a maximum of 61
antennas may contribute to the phased sum.

Optionally, it is possible to retain the last-applied phases in the
TFB phase registers rather than clearing them at the start of a
subscan sequence for an observation.  This allows a ``passive'' phasing
mode whereby a relatively bright calibrator
source located within a few degrees of the fainter target
is used to phase up the array  prior to slewing to the target
and observing it without active phasing. Optimal cycle times are
currently being
explored, but are expected to be of order one minute (see e.g.,
Holdaway 1997; Carilli et al. 1999). 
This mode of operation has been successfully tested during APP CSV (Section~\ref{CSV}) for
sources and calibrators with angular separations up to several degrees
and will enable use of the APS on much fainter science targets than is
presently possible. Passive phasing is expected to be
offered for science observing in future ALMA Cycles (Section~\ref{future}).

\subsubsection{The ``Slow'' Phasing Loop}
\label{sec:slow-loop}
Data are delivered to TelCal in the form of channel averages at the end
of each correlator subscan.  The number of such channel averages in
each of the four 2~GHz basebands is programmable,
but APS operations in Cycles 4 and 5 use eight channel averages,
each spanning 250~MHz (see Section~\ref{sec:issues} for an
explanation of this choice.)
Sufficient time
resolution on these channel averages is provided to allow data from 
the start of the correlator subscan to be excluded without much
loss of accuracy in the phase calculation. As shown in
Figure~\ref{fig:phasing-scan}, the phasing correction is calculated
at the end of one subscan and applied at the beginning of the next.
In typical APS
operations during ALMA Cycle~4, there was a new subscan every 20 seconds, of
which 12 seconds could be used to determine the phasing solution.  This timing constraint
is dictated by the network bandwidth available to transfer the data.

The phasing calculations follow a
procedure analogous to the 
standard self-calibration technique commonly implemented in radio interferometry
applications (Pearson \& Readhead 1984; Cornwell \& Fomalont 1989).
For an interferometer of $N_{\rm A}$ elements,
a phase $\phi_{\rm obs} (t)$ can be measured as a function of time $t$ on the
baselines between any two
antennas $i$ and $j$.
The observed baseline phases are the sum of the structure phase of the observed
source, $\phi_{\rm mod}$, the instrumental phase,
$\phi_{\rm ins}$, the atmospheric phase, $\phi_{\rm atm}$, and a thermal noise
term, $\phi_{\rm n}$, specific to each baseline:

\begin{equation}
 \phi_{\rm obs} (t) = \phi_{\rm mod} (t) + \phi_{\rm ins}(t)
+\phi_{\rm atm}(t) + \phi_{\rm n} (t)  
\end{equation}

\noindent The structure phases (which have to be known with sufficient accuracy from a
model) and the instrumental and atmospheric phases have to be removed
from the observed baseline phases, which can subsequently be added coherently.
In the initial APS implementation (used during ALMA Cycle~4), the structure phases are assumed to be
zero---i.e., the target source is assumed to be point-like on the
scales sampled by the ALMA array baselines. 
This is a good approximation for most VLBI target sources, 
but the APS design allows the option for inclusion  of more complex
models in future Cycles.

The instrumental, atmospheric, and thermal noise phases are tied to the antennas
and do not need to be separated. $N_{\rm A}$ atmospheric
phases then need to be determined from the $N_{\rm A}(N_{\rm A}- 1)$
complex visibilities between
pairs of
antennas.  For arrays with $N_{\rm A}>2$ 
the problem is thus overdetermined. 
By selecting one reference antenna (in general, a reliable 12~m antenna
located near the array center) and assigning it zero phase, 
a least-squares minimization allows estimation of the
remaining $N_{\rm A}-1$ antenna phases, $\psi_i$.  Specifically, since the
baseline phase measurement $\phi_{ij}$ between antennas (index $i$ and $j$)
may be expressed in terms of the unknown per-antenna
phases as $\psi_i - \psi_j$, we can form the $\chi^2$ statistic as

\begin{equation}
    \chi^2 ~ = ~ \sum_{i \not= j}
    \frac{[\phi_{ij} - (\psi_i - \psi_j)]^2}{ \sigma_{ij}^2 }.
\end{equation}

\noindent The minimum is found where $\partial \chi^2 / \partial \psi_i ~ = ~ 0$
for every $i$, and this provides a linear system of equations to solve.
Such a solver was present in TelCal prior to the initiation of
the APP.
The antenna-based noise $\psi_{\rm n}$ is reduced by a factor $\sqrt{N_{\rm A}}$
from the baseline-based noise $\phi_{\rm n} (t)$.
The resulting antenna phases,  $\psi_{i}$,
are then provided to the
\texttt{PhasingController}, and their negatives are taken as 
the corrections
to be applied in the TFB phase registers to achieve the optimal
station phases. The APP also implemented an
additional algorithm to resolve phase lobe ambiguities that result if
a phase change close to 2$\pi$ occurs (see e.g., Fomalont \& Perley
1999). The algorithm was based on code originally developed by J. D. Romney (private communication).

A number of notions for determining the quality of the solution on
a per-antenna basis were considered to allow
the \texttt{PhasingController}
to remove problematic antennas (nominally with human supervision).
To this end, the solver reports a normalized ``quality'' (in the range
[0,1], with ``1'' representing the highest quality) 
and also estimates a measure of phasing efficiency by comparing the
visibility amplitude of the sum and comparison antennas relative to
reference and comparison antennas.  
The \texttt{PhasingController} also
has the capability to automatically remove antennas, although this
feature has not yet
been needed in practice.

\subsubsection{The ``Fast'' Phasing Loop}
\label{sec:fast-loop}
Each 12~m ALMA antenna\footnote{The 7~m ACA dishes do not have
  WVR units installed and there are currently no plans to so equip
  them.}  
is equipped with a WVR that monitors a
spectral line of H$_{2}$O.
The resulting data may be used to estimate the component of the delay due to water
vapor for that antenna (Nikolic et al. 2013).  These data are provided
and processed approximately
once per second so that the CDP spectral processors may make adjustments
to the baseline phases due to the variable differences in path lengths to
the antennas at the end of each baseline.  Alternatively, the data may be
saved in the ASDM file so that analogous corrections can be applied
after the observation, if desired.

\begin{figure}
\centering
\hspace{-0.5in}
\scalebox{0.45}{\rotatebox{0}{\includegraphics{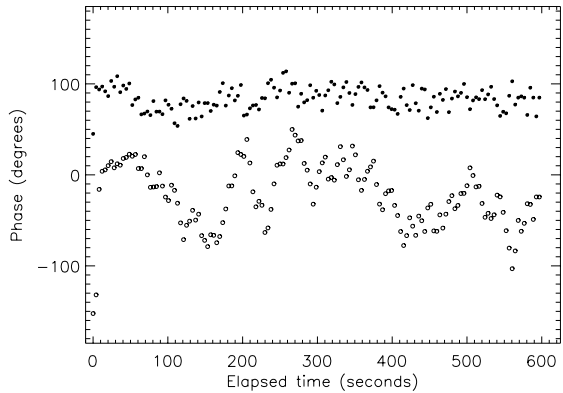}}}
\vspace{-6.0cm}
\caption{Illustration of the effect of the fast phasing loop on phase
  stability during a 5-minute observation of the quasar 3C~279 in
  Band~6. The PWV was $\sim$1.3~mm.
Data for a single 280~m ALMA baseline in one polarization (YY) are
  shown. The open circles show phase as a function of time for 
frequency-averaged data from one 2~GHz quadrant where no phasing
  corrections were applied. The filled dots show data from a second
 quadrant where fast phasing loop corrections (Section~\ref{sec:fast-loop})
were applied. The phase offset between the quadrants is
arbitrary. The mean absolute deviation of the corrected phases is a
factor of 3 lower than that of the uncorrected phases. }
\label{fig:fastloop}
\end{figure}

The WVR data are also usable by the phasing system.  
A software agent was created within the CDP cluster
to translate the WVR data into a series of phase adjustments (approximately once
per second) which are provided to the
\texttt{PhasingController} to pass on to the CCC for application
at the TFB registers.  This ``fast'' loop is ``open'' in the sense that it 
does not provide feedback directly to the CDP agent generating
the phasing corrections (see Section~\ref{sec:slow-loop}).  
In normal use, however, the TelCal phasing
engine is correcting phases that have been modified by WVR corrections,
so in that sense the loop is closed on the slower timescale.

While it is envisioned that the fast loop normally will operate in tandem
with the slow loop (Section~\ref{sec:slow-loop}) 
at times of high water vapor variability, it can also
be operated independently, either for testing, or during
passively phased observations (see Section~\ref{approach}).  
An example of the improvement in phase
stability as a result of use of the fast phasing loop alone (no slow
loop corrections) is shown in Figure~\ref{fig:fastloop}. The data are
from a Band~6 observation in conditions with PWV
$\sim$1.3~mm. Fast-loop corrections were applied to data from one
correlator quadrant, while another quadrant received neither fast nor
slow phasing corrections. The absolute mean deviation of the corrected
phases is a factor of three lower compared with the uncorrected
phases. However,  since the fast corrections are applied with
some latency (about one second), in rapidly varying conditions they
may {\it add} phase noise rather than remove delays.  And in excellent
conditions without much variability, the fast loop is not needed, as little coherence
is lost on the timescale of the slow cycle, hence the corrections
merely add noise.  Because of this, the fast
loop was not used during the initial APS operations in ALMA
Cycle~4. However, it is expected that it may be used in future Cycles
during times of high precipitable water vapor (PWV) and/or if the phasing system is operated at
higher frequencies (see Section~\ref{future}).

\subsubsection{Delay Correction}
\label{sec:issues}
In the ALMA delay system, the geometric
delay is partitioned into a number of pieces for application in several
different parts of the system.  Small, sub-sample delay adjustments  ($<$250~ps)
are made by adjusting the phase of the  digital sampling clock (DGCK)
at each antenna (i.e., 250~ps/16 = 15.625~ps); 
larger adjustments (multiples of 250~ps, or 1 sample) are made
in the station cards of the correlator by shifting the read points of samples queued for
correlation. All of these coarse and fine delay tracking steps are
synchronized in the ALMA system. The smallest adjustments ($<$15.625~ps) are
residuals which cannot be applied in hardware and so are applied
in the spectral processing software of the CDP nodes. 
The latter is also a suitable place to apply the
(relatively stable) baseband delays (which result from differing signal paths
in filters, cables, etc.). Although these are not nearly as
small as the residual delays, they require functionally the same type
of correction (a phase rotation). This correction is normally made on
the data presented to TelCal, but for the APS they need to be applied on the uncorrected
signals in the TFBs, creating the need for an alternative approach.  

The solution implemented during ALMA Cycle~4 was to
disable the baseband delay correction and instead compute and apply the needed
correction as part of the phasing corrections.  By subdividing the
baseband into a set of channel averages with sufficient signal
to robustly calculate phases, the baseband delay correction is then
effectively removed through a delay-like set of phase adjustments across
the channels.  The ALMA system was also modified to support up
to 32 channel averages (one per TFB). The number actually used may 
be tuned according to the source strength, although the default for
Cycle~4 and future Cycle~5 APS observations is 8 channel averages per baseband.
This results in a minor correlation loss (caused by the small residual
delay within each channel-average window)
and necessitates a special processing approach for the ALMA interferometry 
data taken while using the VOM, 
but is otherwise acceptable for the primary
goal of enabling VLBI.
There are plans to address this correction more rigorously in a future
ALMA software release.  The solution will include modification of the phasing logic performed
in TelCal to take the baseband delays into account in the calculation of
the phasing solution. 

\subsubsection{VEX2VOM and the Observing Script\protect\label{vex2vom}}
VLBI observations are specified through a VEX file which captures
the scheduler's intents in a format that can be parsed at each participating
observatory to correctly execute the observations.
At ALMA, observations are defined through a series
of Scheduling Blocks (SBs) associated with a particular observing
project. For the VOM  to function at ALMA  (see Section~\ref{VLBIcontroller}), 
an intermediate agent, VEX2VOM, was needed to marry the intents
(most importantly, the schedule) contained in the VEX file with the
ALMA-specific project specifications
as expressed in the SB for subsequent use by a Python observing
script called \texttt{StandardVLBI.py}. For historical reasons, this
is sometimes referred to as an ``SSR script''.

VEX2VOM checks the VEX file for viability and embeds important pieces of static
information into the SB in the form of ``expert'' parameters.
\texttt{StandardVLBI.py} can then read these parameters from the SB and make all
the necessary run-time decisions prior to executing the observation.  

VLBI observations are a
specialized version of the normal interferometric observations, and \texttt{StandardVLBI.py} and
\texttt{StandardInterferometry.py} share many lower-level
objects. In particular, all
of the standard ALMA Observatory calibrations are executed in exactly the
same way as they would be for normal interferometry. However, there
are enough differences in detail that the two scripts look rather different at the
top-level.  

It is usual for VEX files to need to accommodate the differing capabilities
of an inhomogeneous array of VLBI stations.  Typically the schedule includes sizable
gaps to allow time for either slews to new targets or for observatory-specific
calibrations.  The ALMA dishes slew relatively quickly, so an important
design consideration of \texttt{StandardVLBI.py} was to allow ALMA-specific
calibrations to be carried out during the gaps between VLBI scans.

One other attribute of the VOM worth noting is that it is not
compatible with the use of subarrays, which are now commonly used
during standard interferometry observations with the BL correlator at ALMA in order to enable
parallel activities using different subsets of antennas. The most
significant obstacle for doing VLBI with subarrays is that the phasing application protocol has large
overheads in terms of the time required to deliver data to the
Controller Area Network (CAN) bus
compared with other operations, hence it tends to ``lock out'' other
use of the same CAN bus. Conversely, other CAN bus activity (e.g.,
from unrelated observations) has the potential to block the
application of phasing data.


\section{VLBI Correlation}
\label{sec:korrel}
Correlation of VLBI data from ALMA is performed after the
observations at either of two specialized VLBI correlators: one
at the Max-Planck-Institut f\"ur Radioastronomie
in Bonn, Germany, and the other at the MIT Haystack Observatory
in Westford, MA, USA.  Because of the large
data volumes involved, electronic transfer of VLBI data from ALMA to
the correlation sites is not currently
feasible, and instead the recording media (Section~\ref{mark6}) are directly shipped. 

The design of the APS calls for the resulting VLBI data to be
correlated in as standard a manner as possible.
The designated correlators both currently use
the DiFX software correlator (Deller et al 2007, 2011), which has
reached version 2.5 at the time of this writing. The correlation
hardware at both of these facilities is current generation
Intel-based servers with conventional high-speed networking fabric to allow
the large volume of data made possible with ALMA observations to be
correlated in a relatively short amount of time. 

\subsection{Zoom Mode\protect\label{zoom}}
VLBI backends at most current observatories, including the VLBA, GMVA, and several stations of the EHT, 
channelize data into sub-bands whose frequency widths are based on powers of
two (i.e., $2^{n}$~MHz where $n$ is an integer). 
In contrast, the correlator configuration used for APS
operations partitions ALMA's usable bandwidth within each of the four
basebands 
into thirty-two 62.5~MHz channels, each being processed at the 125~MHz
clock rate of the ALMA correlator.
This sample rate is not commensurate with the standard rates used at
other VLBI stations.

The original APS design proposed to accommodate these different sampling
rates through the implementation of multiple ($\ge$16) 
ALMA spectral windows across each baseband, each tuned with the DC
edge matching those of the 32~MHz VLBI channels at other sites, and 
with 2.5~MHz gaps between the ALMA spectral windows. However, the ALMA
BL correlator does not presently
support the use of multiple ($>$2) spectral
windows within a single 2~GHz band.  Moreover, the tuning is only
possible to within 31~kHz resolution (due to finite frequency
resolution in the TFB digital LO), which creates additional problems.
Fortunately, DiFX possesses a so-called ``zoom'' mode, which 
was originally intended to limit data to interesting parts of the
spectrum (i.e., spectral lines). This zoom mode turns out to be an ideal solution for
the correlation of ALMA data,
as the non-overlapping portions of the channel can be assigned to
zoom bands and then correlated and analyzed as if that had been
the original channelization.  The functionality and performance of the
zoom mode were extensively tested using simulated data, followed by
testing with real data during CSV (Section~\ref{CSV}).

This approach is found to be robust, but 
leads to some small loss of bandwidth (1~\%) as the incommensurate frequencies
must still be made to line up (for cross-multiplication and accumulation in
the spectral domain) and must further match additional constraints imposed
by DiFX for its internal processing.  Additionally, some existing analysis
tools are not expecting unusually-sized (58.59375~MHz) bandwidths.
For this reason, 58~MHz channels are presently adopted; the 1~\% loss
incurred is
comparable to other compromises within working VLBI systems available today.


\section{PolConvert\protect\label{polconvert}}
The ALMA receivers observe in a linear polarization basis. The main
reason for this is to provide a high polarization purity (i.e., a low
polarization leakage between polarizers) at large fractional bandwidths
(e.g., Rudolph et al. 2007). However, most VLBI stations record in a
circular polarization basis, mainly to simplify the parallactic-angle
correction, which is reduced to a phase between the two polarizer
signals (right circularly polarized or ``RCP'' and left circularly
polarized or ``LCP''). This simplification of the parallactic-angle
effects has the consequence of higher polarization leakage, usually
related to the use of quarter-wave plates, especially in observations
with large fractional bandwidths. The
``mixed'' polarization products resulting from ALMA's dual orthogonal linear
polarization data and the dual circular products recorded at other
VLBI sites are handled at the correlator sites
(Section~\ref{sec:korrel}) during the
post-correlation stage.

Several options were considered by the APP for the
adaptation of the ALMA linear polarizations into a circular basis for
VLBI. These included
applying the conversion to the raw streams at the VLBI backend,
computing it at the correlation stage, and/or using a grid on a reference
antenna for an estimate of the phase between the linear polarizers.
Given that the main goal was to minimize the instrumental polarization
effects during the conversion, the final chosen strategy was to apply a
{\it post-correlation} conversion; i.e., to correlate the
linear polarization signals from the APS and the circular polarization
signals from the rest of the VLBI stations. Such a cross-correlation
produces VLBI fringes in a so-called {\it mixed-polarization} basis,
which can then be converted into a pure circular basis using an algorithm
based on hybrid matrices in the frame of the Radio Interferometer
Measurement Equation (e.g., Smirnov 2011). Details about
this algorithm and the specification of these 
matrices are given in Mart\'\i-Vidal et al. (2016).

\begin{figure}
\scalebox{0.35}{\rotatebox{0}{\includegraphics{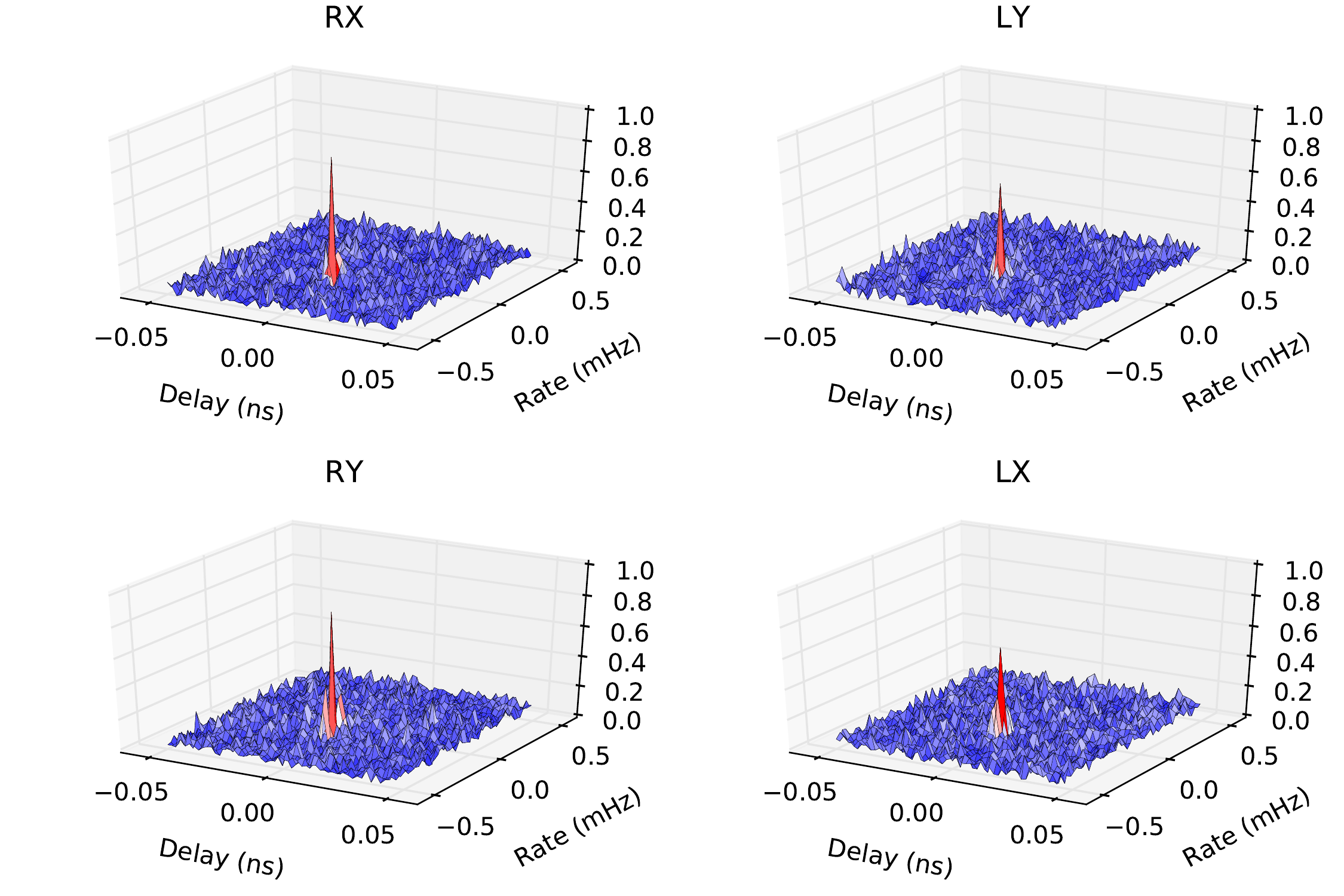}}}
\scalebox{0.35}{\rotatebox{0}{\includegraphics{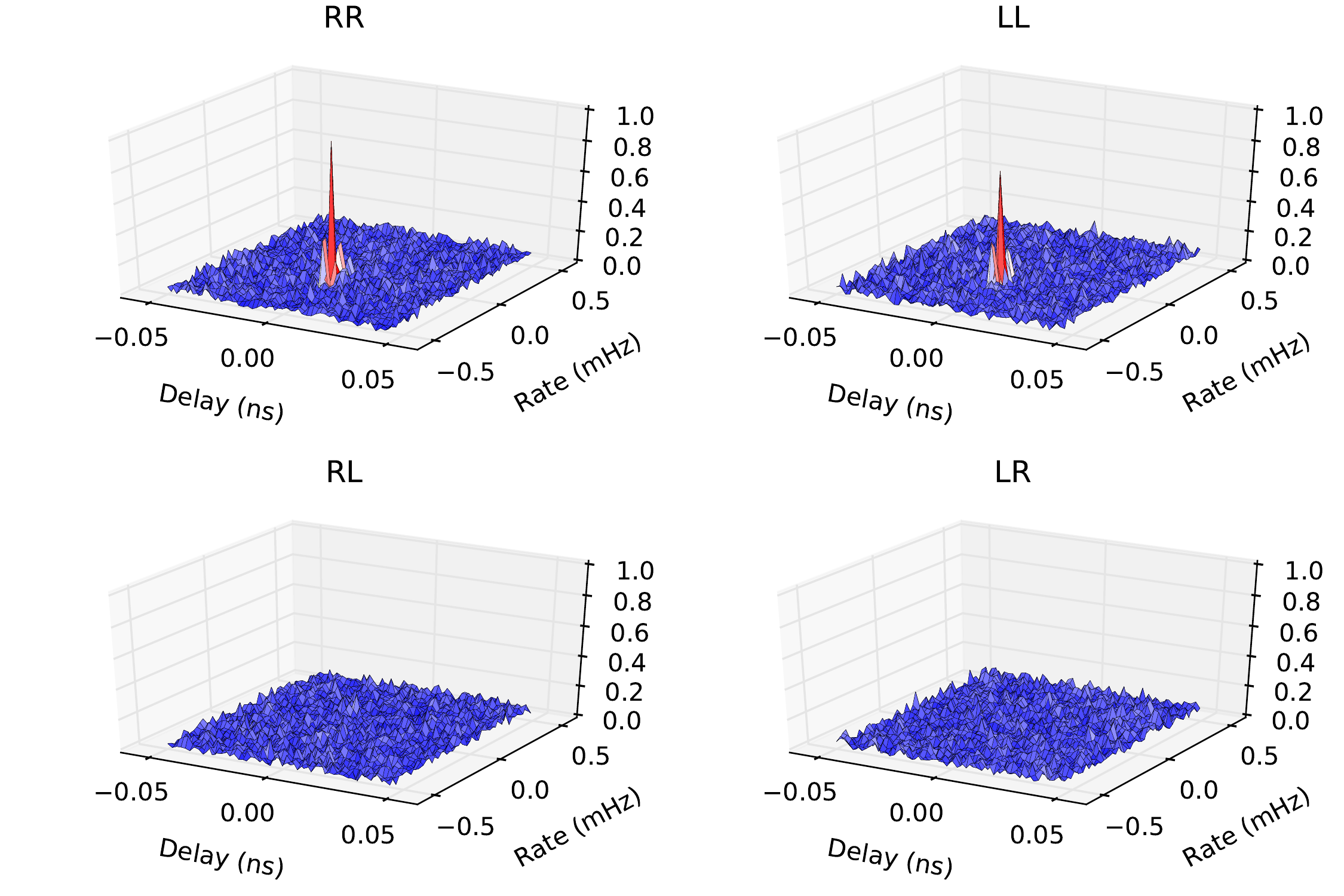}}}
\caption{Delay-rate-fringe amplitude plots on a baseline between ALMA and the LMT
  during an APP CSV Band~6 observation of the quasar PKS J1512$-$0905 in 2016
  April. Panels in the top two rows show the ``mixed''
  linear (X and Y) and circular (R and L) polarization products. The
  bottom two rows show the data after processing with PolConvert to
  translate ALMA's linearly polarized data to circular (see
  Section~\ref{polconvert}). The boost in amplitude of the RR and LL
  fringes  compared with RX and LY, and 
absence of signal in the cross hands of
  the converted data (RL and LR) demonstrate that the conversion is working nominally.}
\label{fig:polconvert}
\end{figure}

There are three main reasons to prefer a post-correlation conversion for
the ALMA signal polarization. First, the later this conversion
is applied to the data, the more reversible the process is. This means that
there is the possibility to perform a refined conversion on near-final data
products, hence minimizing the use of resources. Second, the
hardware required for implementation is minimized, since there is no need to
apply the conversion either at the antenna frontends or at the
VLBI backend; indeed, our strategy resulted in a cheaper and faster
implementation of the APS. Third, an off-line conversion
allows us to perform a full analysis of the ALMA data, in order to find
the best estimates of the pre-conversion correction gains prior to the
polarization conversion (Mart\'\i-Vidal et al. 2016).

The process of polarization conversion can be divided into two main parts.
First, the visibilities among the ALMA antennas (computed by the ALMA
correlator, simultaneous with the VLBI observations) are calibrated
using ALMA-specific algorithms for full-polarization data reduction.
Within the ALMA organization, this process is known as ``Quality Assurance''
of level 2 (QA2). The calibration tables derived in the QA2 stage are subsequently
sent to the VLBI correlators (Section~\ref{sec:korrel}). These tables are
then used by an APS-specific software suite known as PolConvert,
which is run at the correlator computers and applies the polarization
conversion directly to the VLBI visibilities.

In Mart\'\i-Vidal et al. (2016), there is a detailed description of how the
ALMA-QA2 calibration tables are applied to the VLBI data. Working out
their Eq.~18, we can re-write the effect of any residual
cross-polarization gain ratio, $G_{x/y}$, on the ``PolConverted''
visibilities in the form:

\begin{equation}
V_{\odot\odot} = \left(\begin{array}{cc} 1 & D \\ D & 1
\end{array}\right)V^{\mathrm{true}}_{\odot\odot},
\end{equation}

\noindent where $V_{\odot\odot}$ is the visibility matrix computed by
PolConvert, $V^{\mathrm{true}}_{\odot\odot}$ is the true visibility
matrix (i.e., free from instrumental effects), and $D = (1 - G_{x/y})/(1
+ G_{x/y})$ is an instrumental leakage factor, which depends on the
residual cross-gain ratio between the linear polarizers (i.e., $G_{x/y}$).

According to the specifications devised by the APP, the $D$ factor should be
lower than 3~\% in absolute value, which translates into a $G_{x/y}$ of
less than $\sim$5~\% in amplitude and $\sim$5~degrees in phase. These
figures fall within the requirements of ordinary ALMA full-polarization
observations (i.e., a 0.1~\% sensitivity in polarization, or
better; Nagai et al. 2016). 
Therefore, the requirement for a post-conversion
polarization leakage below 3~\% is easily met after the QA2 calibration
is performed. 

Sample delay-rate-fringe amplitude plots before and
after running PolConvert are shown in Figure~\ref{fig:polconvert} for
a Band~6 observation of the quasar PKS J1512$-$0905 on a baseline between
ALMA and the LMT. These data were obtained as part of APP CSV
(Section~\ref{CSV}). It is seen that the initial mixed polarization
correlations RX and LY have lower amplitudes than the RR and LL
products produced after running PolConvert. Following PolConvert, the
cross-terms RL and LR also show negligible signal, implying the
conversion is working nominally.


\section{Performance of  the Phasing System\protect\label{CSV}}
\subsection{Commissioning and Science Verification (CSV)}
CSV for the APP was carried
out over a two-year period from January 2015 to January 2017. CSV activities
used on-sky testing to validate the fully integrated components of the APS,
including hardware, software, observing modes, recording modes, and
correlation procedures. The tests that were carried out
were designed to demonstrate that
the APS met its formal design requirements,
that the APS produces robust, scientifically
valid data whose characteristics are well understood and documented,
and that ALMA is capable of operating as a fully
functional VLBI station. Presently these attributes continue to be checked by
periodic regression testing of the APS.
Details from the various APP CSV campaigns and supplementary test
observations have been described in a
series of ALMA Technical Notes (Matthews \& Crew 2015a, b, c;
Matthews, Crew, \& Fish 2017). 

\subsection{Phasing Efficiency\protect\label{efficiency}}
As noted in Section~\ref{background}, a perfectly phased array of
$N_{\rm A}$ antennas  is
equivalent  to a monolithic aperture with $N_{\rm A}$ times the effective area of
one of the individual elements, $A_{\rm eff}$. However, in practice, any real phased
array will suffer from efficiency losses (i.e., loss of effective
collecting area) caused by a combination of
factors. These losses may be collectively 
characterized in terms of a ``phasing
efficiency'', $\eta_{\rm p}$, where $\eta_{\rm p}$=1 corresponds to perfect efficiency.
Routine characterization of the phasing efficiency allows determining 
whether the system consistently 
meets theoretical expectations, and is crucial for  
enabling accurate absolute calibration of VLBI
experiments. In addition, such a measure serves as a useful metric for
comparing performance of the system under
various conditions (e.g., different weather, array configurations, or observing frequencies).

The gain of an antenna may be defined as
$A_{\rm eff}/(2 k)$ where  $k$
is the Boltzmann constant (TMSIII). If the effective area of a single ALMA
antenna is $A_{\rm 12m}$, then the gain of a phased array of $N_{\rm A}$ such
antennas may be written as: $N_{\rm A}A_{\rm 12m}/(2 k)$.

If we now correlate the phased sum signal $V_{\rm sum}$ (Eq.~1)
with the signal $V_{\rm c}$ from another
identical ``comparison'' antenna in the array that is {\it not} part of the
phased sum, the resulting product can be written as

\begin{equation}
\langle V_{\rm sum}V_{\rm c}\rangle = \left(\frac{V^{2}_{0}}{2}\right){\rm
  cos}(2\pi\nu t)
\end{equation}

\noindent where $\nu$ is the observing frequency, $t$ is the
integration period, and $(V^{2}_{0}/2)$ is the amplitude of the
correlated signal 
(e.g., Condon \& Ransom 2016). 
Because $V_{\rm sum}$ and $V_{\rm c}$ are both 
proportional to the electric field of the
source being observed, multiplied by the gains of the sum antenna and
comparison antenna, respectively, this
implies that the amplitude term in Eq.~6 is in turn
proportional to $\sqrt{(A_{\rm 12m})(N_{\rm A}A_{\rm 12m})}$.
Thus for an ideal phased array, 
the correlated amplitude is expected to grow as the
{\it square root} of the number of antennas used to construct the sum. 

The {\it phasing efficiency} of the array, $\eta_{\rm p}$, may then be
parametrized as 

\begin{equation}
\eta_{\rm p} 
= \frac{\langle V_{\rm sum}V_{\rm c}\rangle }{\sqrt{N_{\rm A}}\langle
  V_{i}V_{\rm c}\rangle }
\end{equation}

\noindent where the $\langle
V_{i}V_{\rm c}\rangle $ term in the denominator denotes
the mean of the correlated amplitudes between each of the $N_{\rm A}$ antennas
used to form the phased sum and the unphased comparison antenna.
(TelCal computes this ratio for a designated comparison antenna and provides the
result within the ASDM file).
Throughout the APP CSV, a variety of tests were used to gauge the overall 
efficiency of the APS empirically and to isolate the sources and
magnitudes of various losses. 

One means to
characterize the efficiency of the phasing system 
is the execution of 
``step scan'' sequences (Primiani et al. 2011)  
on bright, compact sources. During such an observation,
antennas are sequentially added to, or subtracted from the phased
array, permitting an evaluation of how correlated amplitude (and
hence the
phasing efficiency) scales with the number of antennas in the phased
array.  
During a step scan test, one or more array antennas are 
designated as comparison antennas and are never phase-corrected or 
included in the phased sum. 

\begin{figure}
\hspace{-0.5in}
\scalebox{0.5}{\rotatebox{0}{\includegraphics{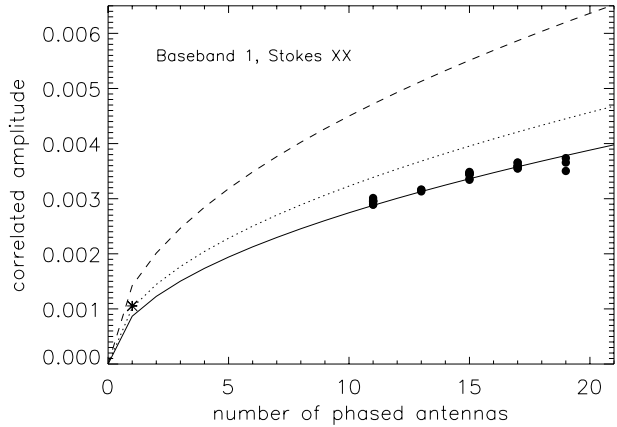}}}
\vspace{-6.0cm}
\caption{Results of a ``step scan'' phasing efficiency 
test of the APS performed on 2015
  March 31 in Band~6.  
During a $\sim$10-minute observation, pairs of antennas
  were sequentially added to or 
removed from the phased array. The filled dots indicate 
correlated amplitude as a function of the number of phased antennas 
for baselines between the phased sum and an unphased comparison
antenna. Each point is based on 16~seconds of data; results
for only one of
  the eight baseband/polarization combinations are shown. The asterisk
designates the correlated amplitude  prior to
  phase-up of the array (i.e., it is comparable to a phased array of
  one antenna). The solid
  line shows a fit to the data (for $N_{\rm A}>1$) 
of the form $A\propto \sqrt{N_{\rm A}}$
  where $N_{\rm A}$ is the number of phased antennas. The dotted
  line is the predicted result based on the $N_{\rm A}=1$ measurement,
  while the dashed line shows the prediction for an ideal phased array. 
The results indicate that the amplitude generally increases
with the number of phased antennas and that the overall phasing
efficiency during the observation was $\sim$61~\%. See Section~\ref{efficiency}
for
details. }
\label{fig:stepscan}
\end{figure}

Several step scan tests were performed using the APS during CSV campaigns. 
Results from one example are presented in
Figure~\ref{fig:stepscan}. 
These results are based on a Band~6
observations of the quasar PKS J0522$-$3627
obtained with the APS 
on 2015 March~31. The weather conditions were relatively stable with
PWV$\sim$1.3~mm.  
There were a total of 22 functioning 12-meter antennas in the
array used for the test, three of which were designated as unphased comparison
antennas. Baseline lengths ranged from 11~m to 313~m. During
the course of a 10-minute observation, pairs of antennas were
randomly added to or removed from the phased sum (hereafter
designated ``APP001'', 
corresponding to the station identification of the phased sum in the
ASDM file),\footnote{Strictly speaking, APP001 is a fictitious pad position
corresponding to the array reference position.}  with the maximum
number of phased antennas being 19. The test was performed independently
in the four basebands, resulting in different values of $N_{\rm A}$ in
each quadrant at any given time. For clarity, we show here only
the results for a single
baseband/polarization combination (polarization XX in baseband 1,
where the center frequency was 214.6~GHz) and for only one of the comparison
antennas (``DA45''). Results for the other basebands and
comparison antennas are comparable.

Figure~\ref{fig:stepscan} shows the correlated amplitude for the
DA45--APP001 baseline as a function of the number of phased
antennas for each 16~s subscan during the step scan test (closed
dots). For each time interval, data from the 8 subbands within the
baseband were averaged to yield a single amplitude value, and
data from the first minute of data (during which the array
was not fully phased) were excluded. 

The solid line plotted on Figure~\ref{fig:stepscan} 
shows the result of a
least squares fit to the measured amplitudes of the form
$A=C\sqrt{N_{\rm A}}$,
where $C$ is an arbitrary scaling factor. The correlated amplitude, $A$, is
seen to
follow the expected $\sqrt{N_{\rm A}}$ dependence for $11\le N_{\rm A}\le 17$,
but it is slightly lower than expected for $N_{\rm A}$=19.
For
comparison, the asterisk plotted on Figure~\ref{fig:stepscan} shows the
correlated amplitude on the DA45-APP001 baseline {\it prior} to
phase-up, i.e., effectively showing the amplitude for a phased array
of one antenna. The dotted curve running through this latter point
illustrates the correlated amplitudes that are predicted if this is adopted
as a benchmark for behavior of the system. The actual
measured values lie $\sim$15~\% below this curve.
Finally, the dashed curve on  Figure~\ref{fig:stepscan} plots the
correlated amplitude expected for a perfectly phased array
($\eta_{\rm p}=1$). We see that the measured amplitudes are $\sim$61~\% of
those expected based on this idealized prediction. 
Unlike the values used
to produce the dotted curve, those used to produce the dashed curve
have not been processed by the summer mechanism. This therefore provides
us a means of disentangling efficiency losses due to factors inherent
to phase-correcting the antennas versus those inherent to the production of the
phased sum signal itself.

Efficiency losses in the APS 
occur at different stages.  First, imperfections in the 
phasing solutions themselves (as a consequence of tropospheric fluctuations, residual delay errors,
source structure, and latency in the
application of the phasing corrections) all increase the RMS
scatter, $\sigma_{\phi}$, in the phases of the antennas used to
form the phased sum. 
The decorrelation resulting from these fluctuations in turn
produces a reduction in the correlated amplitude by a factor
$\epsilon_{0}\approx e^{\frac{-\sigma^{2}_{\phi}}{2}}$ (e.g., Carilli et
al. 1999). For the data set
shown in Figure~\ref{fig:stepscan}, $\sigma_{\phi}$=0.37 rad (21$^{\circ}$), hence
$\epsilon_{0}\sim$0.93. In general, the
atmosphere dominates $\epsilon_{0}$, hence efficiency will be worse at
higher frequencies and in times of poor atmospheric
stability. Nonetheless, in terms of these losses, the APS phasing
engine routinely
meets or exceeds its goal of 90~\% phasing efficiency as specified in the original
functional requirements (see Section~\ref{overview}).

Second, a significant and unavoidable loss of phasing efficiency in the APS 
arises from the 2-bit quantization of the summed
signal, which leads to an amplitude reduction of 12~\% (i.e.,
$\eta_{\rm q}=0.88$; e.g., Kokkeler et al.
2001; Crew 2012).
Third, additional smaller efficiency losses arise from discrete sum scaling
errors (up to 5~\%) and the lack of optimized weighting of the
contributions of the individual antenna signals to the sum.
Ideally, one would want to weight the antenna contributions 
to the sum using either equal power or variance weighting 
(Moran 1989). However, the correlator circuit that receives the antenna data
to be summed  
(provided by one of the so-called VLBI hooks; Baudry et al. 2012) 
does not have access to any additional data and 
has logic space only for a simple sum and not for additional weighting
factors.
Because the ALMA 12~m antennas  are
generally well matched (ALMA Partnership, private communication), 
the weights should be approximately equal and
this type of efficiency loss is not expected to exceed a few per
cent. 

\begin{figure}
\hspace{-0.5in}
\scalebox{0.5}{\rotatebox{0}{\includegraphics{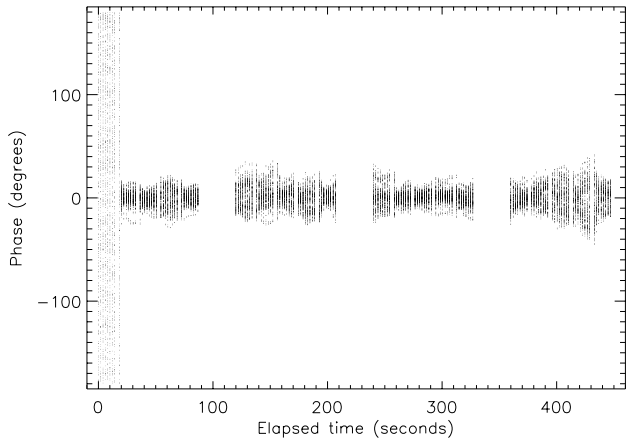}}}
\vspace{-6.6cm}
\caption{Phase as a function of time during a series of four Band~6
  APS scans on
  the $\sim$3~Jy source PKS J1924$-$2914, conducted on 2017 April 10 using an array of 37
phased 12~m antennas. Phases are plotted 
for baselines between the phasing reference antenna
and each of the individual antennas used to form the phased sum. Data for all four ALMA
basebands, in both the XX and YY polarizations, are shown. Phase-up
is achieved approximately 20 seconds into the first scan, after which the mean RMS
phase scatter is $\sim10$~degrees.  }
\label{fig:nicephases}
\end{figure}

Figure~\ref{fig:nicephases} shows results from
another test of the APS, conducted on 2017 April 10 in Band~6 using an array of 37
phased 12~m antennas. The goal was to assess the phasing quality
and stability as a function of time for all basebands and
both polarizations.
In this case, the number of phased antennas was
kept fixed throughout. Four scans were obtained on the quasar PKS J1924$-$2914,
during which the array was actively phased. Based on observations from
the SMA\footnote{The 1~mm flux density of J1924$-$2914 close to the date
  of observation was obtained
  from the SMA calibrator web site:
  http://sma1.sma.hawaii.edu/callist/callist.html. Measurements at
  this site are  obtained with the SMA as part of an ongoing
monitoring program (Gurwell et al
2007). 
The measured source signal strengths are calibrated
against known standards, typically solar system objects. }, the 1~mm source flux
density
near the date of observation was 3.18$\pm$0.16~Jy.
The baseline lengths ranged from 14~m to 400~m and the PWV
was $\sim$0.5~mm.
The figure shows the RMS scatter in the phases as
a function of time for baselines between the phasing reference antenna
and the antennas used to form the phased sum. At start-up, the array was
completely unphased, as evidenced by the large phase scatter. Phase-up
is achieved after approximately 20 seconds, after which the mean RMS
phase scatter is $\sim10$~degrees across all of the four basebands, both
polarizations, implying a decorrelation loss in amplitude of $<2$~\%. 

These and
other test data taken throughout CSV and subsequent regression testing
show that under stable
weather conditions, efficiency losses as a result of imperfections in
the phasing solutions themselves are generally minimal. 
Note that during the
second and fourth scans in the present example, the fast phasing loop was
used in concert with the slow loop. We see that the RMS
scatter is $\sim$30~\% higher during those scans, confirming that these WVR-based
corrections add noise when used under very stable weather 
conditions (see Section~\ref{sec:fast-loop}).

Importantly, the data shown in Figure~\ref{fig:nicephases} have been corrected by
the APS phasing software, but they have not been processed by the
summer, and are therefore not subject to the degradation in the signal caused by
quantization errors and other subsequent losses. To assess these
latter effects,  we may examine the correlated
amplitude on a baseline between the phased sum
signal and an unphased comparison antenna, $a_{S \otimes {\rm c}}$, with
respect to the correlated amplitude seen on a baseline between a
comparison antenna and the phasing reference antenna $a_{{\rm c} \otimes
  R}$. For a perfectly phased array subject to 2-bit quantization
losses, the predicted ratio $a_{S \otimes {\rm c}}/a_{{\rm c} \otimes  R}=
0.88\sqrt{N_{\rm A}}$  where $N_{\rm A}$ is the number of antennas in the
phased sum. Instead, for the data set shown in
Figure~\ref{fig:nicephases}, we find $a_{S \otimes {\rm c}}/a_{{\rm c} \otimes
  R}\approx0.61\sqrt{N_{\rm A}}$, consistent with the results of the step
scan test described above. Similar total efficiency values were also confirmed in
subsequent VLBI observations (see Section~\ref{fringes}). 

Some portion
of the APS efficiency loss can be attributed to the present treatment of
baseband delays (see Section~\ref{sec:issues}). This effect is
expected to be of order a few per cent (Crew 2012). Additional factors
contributing to an efficiency loss of $\sim$20~\% above the
theoretical expectation
remain a topic of ongoing investigation.

\subsection{VLBI Fringes\protect\label{fringes}}
\begin{figure*}
\scalebox{0.6}{\rotatebox{0}{\includegraphics{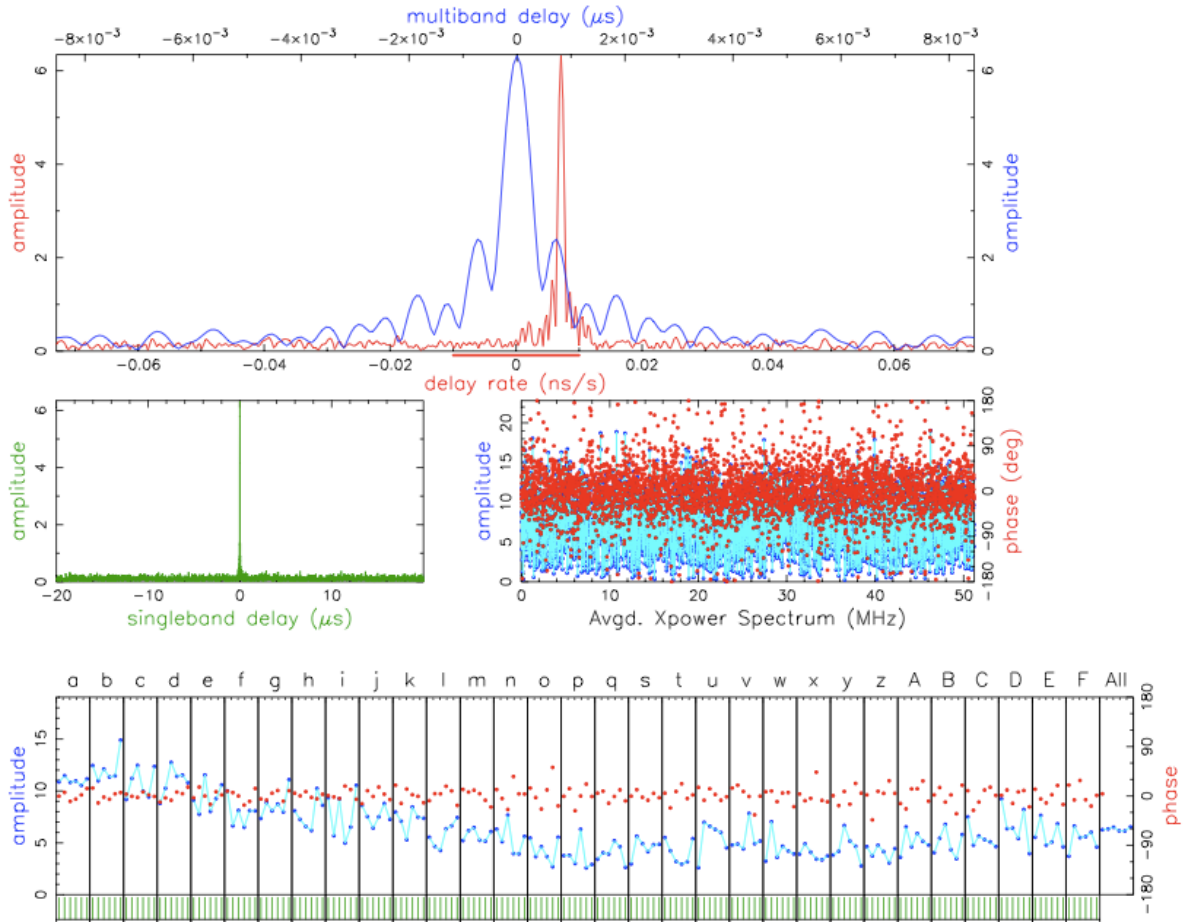}}}
\vspace{0.0cm}
\caption{VLBI fringe detection obtained on a 2.1~km baseline between ALMA and
APEX on 2015 January 13 in Band~6 on the quasar PKS J0522$-$3627. Five seconds
of data are shown. The data are of ``mixed'' polarization (ALMA's
linear X
polarization correlated against APEX's RCP data product).
The  clear peaks 
in the delay rate (red curve, top panel) and in the single-band
delay (green curve, middle-left panel) and multi-band delay (blue
curve, top panel) are consistent with the high SNR of the detection
($\sim$68). The ALMA array was unphased during this observation, hence its
sensitivity was equivalent to that of a single 12~m antenna.
The center-right panel shows amplitude (blue) and phase (red),
averaged over all baseband channels at a resolution of 100~kHz. 
The lower panel shows amplitude (blue) and phase (red) as a 
function of time for each of the thirty-two 51~MHz frequency channels. 
(The wings of the 62.5~MHz ALMA TFB channels were excluded). The
fall-off in amplitude across the band is largely due to the
characteristics of the APEX digital backend, which has better response
at low frequencies. 
 }
\label{fig:APEXfringe}
\end{figure*}

The combined functionality of the various hardware components installed by the
APP to enable VLBI operations at ALMA was 
successfully demonstrated for the first time on 2015 January 13 when
a VLBI fringe on the quasar PKS J0522$-$3627 was detected at a frequency of
214.0~GHz ($\sim$1.3~mm, i.e., Band~6) 
on a 2.1~km baseline between ALMA and the
APEX telescope (Figure~\ref{fig:APEXfringe}).  
Despite the short baseline, both stations used
separate, independent frequency references, hence this constituted a
true VLBI experiment.  
Although the ALMA station was operated as a multi-antenna array during
this test, the array was unphased, and therefore the
equivalent sensitivity was that of a single 12~m antenna. 

The SNR expected for detection of a VLBI fringe between ALMA and
another station may be expressed as:

\begin{equation}
{\rm SNR} = \eta_{\rm p}\eta_{\rm b}S \left[ 
\frac{  \sqrt{n_{\rm pol}\Delta\nu t}}{\sqrt{({\rm SEFD}_{1}\times
        {\rm SEFD}_{2})} } \right]
\end{equation}

\noindent where we assume $\eta_{\rm p}$ is the efficiency of the APS
(see Section~\ref{efficiency}), $\eta_{\rm b}$ is a digital loss factor
equal to 0.88 for 2-bit
correlation of VLBI signals, $S$ is the source flux density in Jy, 
$n_{\rm pol}$ is the number of polarizations, $\Delta\nu$ is the
effective observing bandwidth, $t$ is the integration time in seconds, and
${\rm SEFD}_{1}$ and ${\rm SEFD}_{2}$ are
the system equivalent flux densities (in Jy) of the two respective VLBI
stations. The fringe shown in Figure~\ref{fig:APEXfringe} is based on
5~seconds of data, a single polarization (mixed linear-circular, R-X), and a usable
bandwidth of 1632~MHz (32 channels of 51~MHz). Assuming a source flux
density\footnote{Ibid.} of $\sim$3.6~Jy
and an SEFD of 3600~Jy for both the APEX and ALMA stations, the
predicted SNR is $\sim79\eta_{\rm p}$, while the observed SNR is 68 
(note that these data are uncorrected for the use of mixed polarizations as
described in Section~\ref{polconvert}).
This implies $\eta_{\rm p}\approx 0.86$, consistent with a nominal phasing
efficiency after accounting for 2-bit quantization of the sum signal.
Because the array was not summed in this case,
$\eta_{\rm p}$ is not subject to the additional efficiency losses
described above for the $N_{\rm A}>1$ case.

Several subsequent VLBI fringe tests were performed with ALMA as a phased
array, both in Band~3 and in Band~6 (Matthews \& Crew 2015b, c;
Matthews et al. 2017). These tests included VLBI
sessions of several hours duration during which a full suite of ALMA
and VLBI-specific calibration sources were observed in order to permit
end-to-end testing of PolConvert (see Section~\ref{polconvert}). 
One set of observations  (experiment code
BM452) was performed on 2016 July 10 in conjunction with the
8 stations from the National Radio
Astronomy Observatory's\footnote{The National Radio Astronomy
  Observatory (NRAO) is a facility of the National Science Foundation
  operated under cooperative agreement by Associated Universities, Inc.}
VLBA that are equipped with 3~mm receivers.\footnote{In 2016
  October, operation of the VLBA was transferred from NRAO to the Long Baseline Observatory.} 
Two stations (North Liberty and Pie Town) were
excluded from further analysis owing to poor weather.

\begin{figure*}
\scalebox{0.6}{\rotatebox{0}{\includegraphics{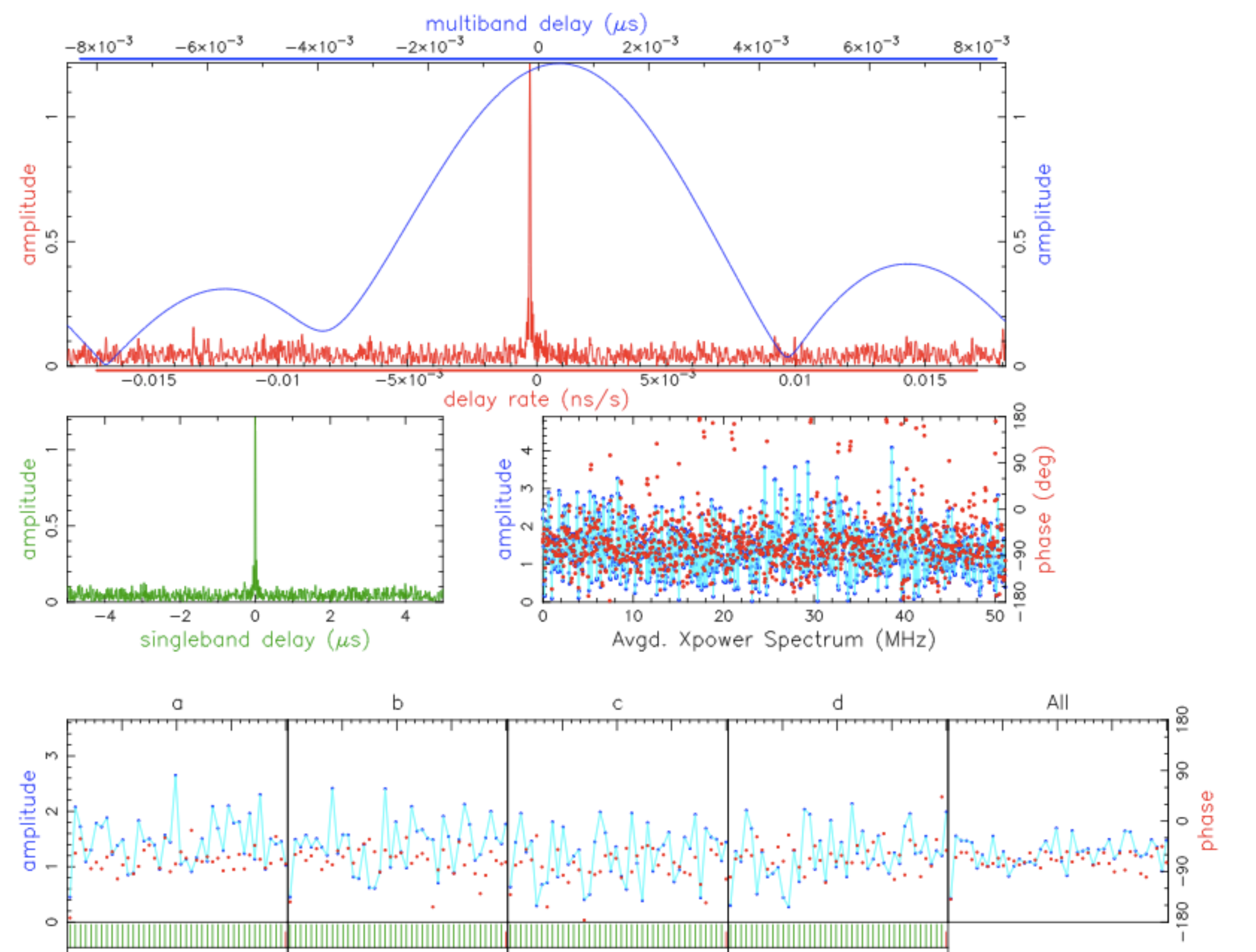}}}
\caption{Fringe plot based on Band~3 scan  of 238~s duration on the target source
PKS J0607$-$0834 on a 6737~km baseline between ALMA and the Fort Davis
VLBA station, obtained on 2016 July 10. 
The ALMA data have been converted from linear to circular
polarization using PolConvert (Section~\ref{polconvert}); the
resulting LL
polarization is shown.  The
SNR of the fringe detection is $\sim$33.
The panels are as described in Figure~\ref{fig:APEXfringe}, with the exception
that the Band~3 data shown here have a smaller total bandwidth (204~MHz, divided into four
51~MHz frequency channels, as shown in the lower panels). 
}
\label{fig:VLBAstation}
\end{figure*}

An example of a fringe on the 6727~km 
baseline between a phased array of 11 ALMA antennas
and the Fort Davis
station of the VLBA is shown in Figure~\ref{fig:VLBAstation}. The usable observing
bandwidth was 204~MHz (four 51~MHz channels per polarization). The
target source was the quasar PKS J0607$-$0834, which had a 3~mm flux density of
$\sim$1~Jy near the time of the observations (see Footnote~4). The
peak source flux density is therefore comparable to the
faintest VLBI science targets permitted during ALMA Cycles~4 and 5. However,
the correlated flux density on the observed baseline is
expected to be approximately four times lower (see Kovalev et
al. 2005). 
Assuming an SEFD for the Fort Davis station of 3500~Jy, an SEFD
for the phased ALMA array\footnote{The estimated SEFD for phased ALMA 
during this test  assumes 11 identical 12~m antennas, each with an aperture efficiency
  of 0.71 and a Band~3 system temperature of 70~K.} of 219~Jy, and an integration time of
238.4~s, the predicted SNR is $\sim$33 after accounting for known
efficiency losses as described above (i.e., $\eta_{\rm p}$=0.61 and 
$\eta_{\rm c}$=0.88) and applying PolConvert
corrections (Section~\ref{polconvert}). Using an optimal segmentation
time of $\sim$6~seconds for the data (to minimize decorrelation losses in
the correlated amplitude), the measured
SNR is 32.5, in excellent agreement with our previous
characterizations of the system.  This result confirms that the end-to-end
phasing efficiency of the APS is routinely $\sim$61~\%, to within an
uncertainty of a few per cent.


\section{Future Enhancements}\protect\label{future}
The implementation of the APS described herein, and offered during
ALMA Cycles~4 and 5, provides the fundamental capabilities needed to
perform VLBI with ALMA. However, extensions and expansions of
the APS are currently being planned for  future
Cycles in order to further broaden and diversify the scientific
applications of the phasing system.

One limitation of the current APS is
the requirement that both the science target  and the associated 
calibrators be point-like and sufficiently bright
on intra-ALMA baselines to
allow the phasing computations to be performed on the target
itself. This restriction will be removed with the introduction of a
``passive''
phasing mode  whereby phase-up is performed on a neighboring
calibration source at an appropriate cadence. This will enable ALMA
VLBI observations on significantly fainter targets, with the flux
density limited by
SNR considerations on a given VLBI baseline, rather than the need to
phase up on the target of interest.

While the APS has so far been commissioned for use in ALMA Bands~3 and
6 only,
there are no technical obstacles to using the system in other bands,
provided that weather conditions are suitable for
phasing. Implementation of the APS in the existing Band~7 (0.7~mm), the pending
ALMA Band~1 (7~mm), and a possible future Band~2 (4~mm) is expected to be straightforward and require
primarily a modest amount of on-sky testing to determine the optimal phasing
parameters  and the optimal range of array baseline lengths for  use at
these different wavebands. Band~7 operation of the APS is of particular
interest for black hole studies, since the blurring effects of
interstellar scattering will be minimized, thereby improving the ability to
reconstruct images of the
source (Johnson \& Gwinn 2015; Johnson 2016). Use of the APS at still
higher frequencies is also possible in principle, although it is
unlikely to be advantageous owing to the increased atmospheric opacity
at these wavelengths and the lack of other VLBI sites capable of
commensurate operation. At the lower end of ALMA's frequency range of operation, Band~1
will provide frequency overlap with the Karl G. Jansky VLA,
and offer an exquisitely sensitive baseline between ALMA and the
phased VLA. Band~1 is also expected to be especially useful for 
pulsar searches (see below), as it provides a good
trade-off between sensitivity and the minimization of interstellar
scattering effects (e.g., Fish et al. 2013). 

The ALMA frequency bands contain numerous astrophysically important
spectral lines, and the observation of certain of these lines can benefit
from the extraordinarily high sensitivity and/or angular resolution
afforded by a phased ALMA (Fish et al. 2012, Tilanus et
al. 2013). These include non-thermal molecular (maser) lines observable in emission
(e.g., SiO, H$_{2}$O, CH$_{3}$OH, and HCN; 
Boboltz 2005; Wootten 2007) and weaker thermal lines 
detectable in absorption. 

Targets for maser
observations include evolved stars, young stellar objects, and the
circumnuclear disks of galaxies, where maser studies with the highest
achievable spatial resolution provide unique
information on physical conditions, kinematics, and magnetic fields (e.g., Richards
2012; P\'erez-S\'anchez \& Vlemmings
2012; Humphreys et al. 2016). Meanwhile, observations of  
absorption lines of HCN, HCO$^{+}$, and other molecules with phased
ALMA will be
possible against background quasars in active galaxies, providing
spatial information on the distribution of molecular gas (e.g.,
Carilli et al. 2000; Muller \& Gu\'elin 2008), enabling
detailed studies of molecular and isotopic abundance ratios as a
function of redshift (e.g., Muller et al. 2011), and providing 
constraints on possible variations of
fundamental constants over cosmological times (e.g., Uzan 2011).  

Enhancements to the APS to enable spectral line mode operation will
include spectral window matching (to allow phase-up on a bright maser
line), optimization of data rate management to allow higher spectral
resolution in the ALMA standard interferometry data (they are
currently spectral averaged by a factor of 8 to decrease data rates),
and the exploration of modified correlation strategies for spectral
line data sets. For example, the fringe-fitting of APS 
continuum data is currently done in the delay-fringe rate plane. But
in the case of a spectral line where power is concentrated in the
frequency domain, 
the delay function becomes broad and fringe-fitting may instead need
to be performed 
in frequency-fringe rate space (Reid 1995). Further, the adopted
correlation approach will need to be robust against the possibility of
multiple peaks
 in the fringe rate resulting from atmospheric turbulence. 

One other mode of APS operation that is expected to be of considerable
interest to the community is use of the phased array as a ``single
dish'' for pulsar searches in the mm domain. While pulsars are
known to be relatively weak at ALMA frequencies, they are nevertheless
expected to be detectable in ALMA bands (Torne et al. 2015;
Mignani et al. 2017), and an important trade-off is that the effects from 
interstellar scattering become less severe at shorter wavelengths. 
High-frequency pulsar observations are
therefore expected to allow identification of pulsars orbiting the Galactic
center that could be used to provide novel tests of General Relativity
(Cordes et al. 2004; Eatough et al. 2015; Psaltis et al. 2016). 
In addition, pulsar studies at mm wavelengths are
expected to elucidate the processes that lead to
observed changes in the spectral index above 30~GHz (e.g., L\"ohmer et
al. 2008). Because pulsars are so faint at mm bands, a
pulsar observing mode will depend on the passive
phase-up mode described above, and will also benefit from
implementation of a source modeler which features that ability to
shift the array phasing center away
from the pointing center. 

\section{Summary}
The APP has led to the successful deployment of  the hardware and software
necessary to operate ALMA as a phased array and as the world's most
sensitive mm VLBI station. These
capabilities enable leveraging the enormous sensitivity of ALMA for
science that requires extraordinary angular resolution, including
studies of black hole physics on event horizon scales. 

The
current APS has been commissioned for use for continuum observations
of bright ($S\ge$0.5~Jy), compact targets
in ALMA Band 3 (3~mm) and
Band~6 (1.3~mm), and the first successful science observations using these
capabilities
were made in conjunction with the GMVA and EHT,
respectively in 2017 April.  
Ongoing and
future enhancements of the APS are expected to continue to expand and
enhance the scientific applications of a phased ALMA in the coming years.


\acknowledgments The APP gratefully acknowledges the
support it received in wide-ranging areas from numerous individuals 
throughout the ALMA organization and at other institutes around the
globe. 
It is nearly impossible to compile a
comprehensive list of those who contributed to the APP's success, but
among them are: H. Alarcon, J. Antongnini, J. Avarias, E. Barrios,
U. Bach, A. Biggs,
T. Beasley, L. Blackburn, J. Blanchard, R. Blundell, A. Bridger,
W. Brisken, M. Caillat, 
R. Cappallo, J. Castillo, C. Chandler, M. Claussen, C. Coldwell, S. Corder, 
F. Cruzat, I. De Gregorio,  A. Deller,
B. Dent,  M. Derome, P. Friberg, S. Fuica, E. Garcia, J.C. Gatica, J. Gil,
C. Goddi, M. Gurwell,
C. Jacques, D. Herrera, R. Hills, A. Hoffstadt, D. Hughes,
J. Ibsen,
H. Johnson,
J. Kern, R. Kneissl, T. Krichbaum, R. Laing, C. Lonsdale, C. Lopez, R. Lucas,
G. Marconi, L. Martinez-Conde, M. McKinnon, R. Lucas,
R. Marson, S. Matsushita,
A. Mioduszewski, G. Narayanan, 
L.-\ang\ Nyman, J. Ogle, G. Ortiz-Le\'on, M. Parra, R. Porcas, R. Primiani, W. Randolph,
A. Remijan, J. Reveco, A. Rogers, R. Rosen, N. Saez, A. Salinas, T. Sawada,
J. Sepulveda, T.-C. Shen, J. SooHoo,
R. Soto, D. Sousa, T. Staig,
M. Straight, S. Takahashi, C. Tapia,
R. Tilanus, M. Titus,  L. Vertatschitsch, E. Villard, J. Weintroub,
N. Whyborn, K. Young, and A. Zensus. The
APP was supported by a Major Research Instrumentation award from
the National Science Foundation (award 1126433), an ALMA North
American Development Augmentation award, ALMA North American Cycle
3 and Cycle 4 Study awards, a Toray Science and Technology Grant
through the Toray Science Foundation of Japan, and a Grant-in-Aid for
Scientific Research on Innovative Areas 
(25120007),
provided by the Japan Society for the Promotion of Science and the 
Ministry of Education, Culture, Sports, Science and Technology (JSPS/MeXT). This work made use
of data from NRAO projects BM434 and BM452.

\end{document}